%% file: main.tex
\documentclass[sigplan,screen]{acmart}

\usepackage[]{hyperref}

\RequirePackage{doi}
\usepackage{url}

\usepackage{breakurl}
\usepackage{stfloats}
\usepackage{multirow}
\usepackage{multicol}

\newcommand\ket[1]{\left|#1\right\rangle}

\usepackage{geometry}

\copyrightyear{2024} 
\acmYear{2024} 
\setcopyright{rightsretained} 
\acmConference[ASPLOS '24]{29th ACM International Conference on Architectural Support for Programming Languages and Operating Systems, Volume 3}{April 27-May 1, 2024}{La Jolla, CA, USA}
\acmBooktitle{29th ACM International Conference on Architectural Support for Programming Languages and Operating Systems, Volume 3 (ASPLOS '24), April 27-May 1, 2024, La Jolla, CA, USA}
\acmDOI{10.1145/3620666.3651372}
\acmISBN{979-8-4007-0386-7/24/04}

\begin{document}

\title{OnePerc: A Randomness-aware Compiler for Photonic Quantum Computing}

\author{Hezi Zhang}
\email{hezi@ucsd.edu}
\affiliation{%
  \institution{University of California,}
  \city{San Diego}
  \country{USA}
}
\author{Jixuan Ruan}
\email{j3ruan@ucsd.edu}
\affiliation{%
  \institution{University of California,}
  \city{San Diego}
  \country{USA}
}

\author{Hassan Shapourian}
\email{hshapour@cisco.com}
\affiliation{%
  \institution{Cisco Quantum Lab}
  \city{San Jose}
  \country{USA}
}
\author{Ramana Rao Kompella}
\email{rkompell@cisco.com}
\affiliation{%
  \institution{Cisco Quantum Lab}
  \city{San Jose}
  \country{USA}
}
\author{Yufei Ding}
\email{yufeiding@ucsd.edu}
\affiliation{%
  \institution{University of California,}
  \city{San Diego}
  \country{USA}
}

\renewcommand{\shortauthors}{H. Zhang, J. Ruan, H. Shapourian, R. Kompella, Y. Ding}

\input{01_abstract}

\maketitle 

\input{02_introduction}
\input{03_background}

\input{04_motivation}

\input{06_tech_1_revised}

\input{06_tech_2_revised}
\input{06_tech_3_revised}

\input{07_evaluation}

\input{08_conclusion}

\input{10_ackn}
\input{ae}

\bibliographystyle{unsrturl}
\bibliography{references}


\end{document}

%% file: 01_abstract.tex
\begin{abstract}

The photonic platform holds great promise for quantum computing. Nevertheless, the intrinsic probabilistic characteristic of its native fusion operations introduces substantial randomness into the computing process, posing significant challenges to achieving scalability and efficiency in program execution. In this paper, we introduce a randomness-aware compilation framework designed to concurrently achieve scalability and efficiency. Our approach leverages an innovative combination of offline and online optimization passes, with a novel intermediate representation serving as a crucial bridge between them. Through a comprehensive evaluation, we demonstrate that this framework significantly outperforms the most efficient baseline compiler in a scalable manner, opening up new possibilities for realizing scalable photonic quantum computing.

\end{abstract}

%% file: 02_introduction.tex
\vspace{1em}
\section{Introduction}
Photonic platform holds great promise for universal quantum computing due to the unique advantages of photonic qubits~\cite{OBrien2009,Bogdanov:17}, including their great scalability, long coherence time and easy integration with quantum networks. 
Besides the experimental demonstration of quantum supremacy on photonic systems~\cite{zhong2020quantum,PhysRevLett.127.180502,Madsen2022}, 
PsiQuantum has proposed their technology roadmap towards one million qubits using silicon photonics~\cite{fbqc, interleaving}.
The potential high clock speed of this approach \cite{interleaving} could make photonic platform advantageous for near-term quantum algorithms.

Photonic quantum computing differs from other platforms such as superconducting~\cite{devoret2013superconducting}, ion trap~\cite{bruzewicz2019trapped} and neutral atoms \cite{saffman2016quantum}, as it is scaled up by a probabilistic operation known as \emph{fusion}~\cite{fbqc}.
Fusion plays a key role of  forming large-scale entanglements between photonic qubits by merging small entangled states into larger ones upon success.
Its probabilistic feature comes intrinsically from the degeneracy in fusions' outputs for different input states \cite{0.95with30}, bringing significant randomness to the computing process.
On the hardware side, improvement of fusion success probability to a high value requires an impractical amount of ancillary resources~\cite{0.78,0.95with30}. On the software side, this randomness is not taken into consideration by existing software infrastructures for the circuit-based model ~\cite{nielsen2001quantum} (e.g., Qiskit~\cite{Qiskit}, Tket~\cite{tket}).
This is because the weak interaction between photons makes it hard to realize 2-qubit gates in the circuit model, but favors a different computing model known as measurement-based quantum computation (MBQC) or one-way quantum computation (1WQC)~\cite{mbqc2009, translation}.

\begin{figure}[h]
    \centering
    \includegraphics[width=0.33\textwidth]{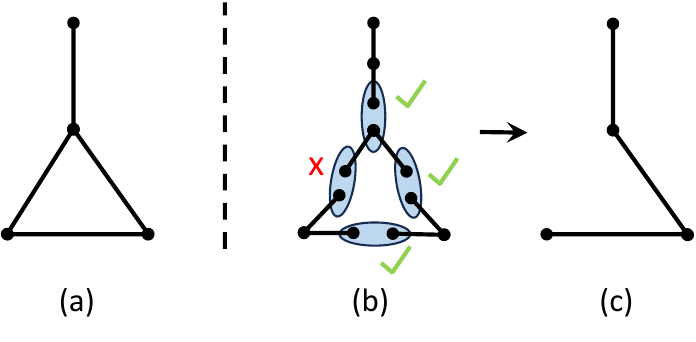}
    \caption{Randomness brought by fusion failures.
    }
    \label{fig:fusion_failures}
\end{figure}

Recently, as an initial effort towards efficient photonic MBQC, a compilation framework OneQ~\cite{OneQ} has been proposed to significantly reduce the depth of compiled programs and the number of required fusions.
However, it overlooks the severe randomness brought by fusion failures, simply assuming that fusions always succeed. 
%
As a compiler, OneQ translates the construction of a program-specific entangled state called \emph{graph state} ~\cite{translation} (Fig.~\ref{fig:fusion_failures}(a)) into a fusion pattern between the small entangled \emph{resource states}~\cite{fbqc}  available on the hardware (Fig.~\ref{fig:fusion_failures}(b)).
However, when fusion failures occur in real-time execution, as illustrated in Fig.~\ref{fig:fusion_failures}(b), the resulting state becomes a random graph state deviating from the target structure. Thus the execution needs to be retried until success, which is non-scalable given that a practical fusion success probability in the near term is merely around 75\%~\cite{0.78, 0.95with30}.
From now on, we will refer to the graph states required by programs 
(e.g., Fig.~\ref{fig:fusion_failures}(a)) 
as \emph{program graph states} and those generated by fusions 
(e.g., Fig.~\ref{fig:fusion_failures}(c)) 
as \emph{physical graph states}.

The objective of this paper is to present a scalable and efficient compilation framework capable of effectively handling fusion failures in the computing process.
This is intuitively a hard problem. Firstly, with a failure rate as high as 25\%, it seems impossible to ensure the formation of any specific entanglement structure among the photonic qubits.
Secondly, failed entanglements such as the disconnected edges in Fig.~\ref{fig:fusion_failures} cannot be recovered since the photons involved in the fusion are completely destroyed by the fusion. Thirdly, the limited lifetime of photons refuses the execution of over-complex algorithms in real-time, as photons are prone to an increasing loss rate with prolonged storage time in fibers~\cite{fiber}.

Fortunately, there are some nice features of fusions and graph states that we can leverage. Firstly, fusion failures are heralded~\cite{linear_optics}, allowing real-time awareness and enabling the incorporation of classical feed-forward~\cite{linear_optics,99feedforward,nonlinear_feedforward} to adjust subsequent operations based on prior fusion outcomes. Secondly, when the fusion success probability exceeds a threshold, the resulting physical graph state contains a long-range-connected component with a high probability. This widely studied phenomenon, known as percolation~\cite{3GHZ,perc_threshold,2D_renorm}, plays a crucial role in providing viable computing resource, inspiring our framework's name, OnePerc (one-way quantum computing based on percolation). Thirdly, a random graph state can be reshaped into any subgraph of it by eliminating the redundant qubits, which can be achieved by measuring them out in $Z$-bases \cite{mbqc2009}.

However,
leveraging these features is highly nontrivial. In addition to the absence of a general fusion strategy to achieve percolation for generic resource states, and the structural mismatch between the high-level program graph state and the low-level random physical graph state, we need to keep in mind the limited time for real-time passes. Specifically, the process associated with the formation of long-range connectivity and the reshaping of the random physical graph state both need to be carried out in real-time, leading to a high demand on their lightweight design. This creates a conflict between the real-time scalability and the program execution efficiency, as a flexible optimization strategy for efficient program execution may require complex algorithms that are not feasible in real-time.

\begin{figure}[t]
    \centering
    \includegraphics[width=0.45\textwidth]{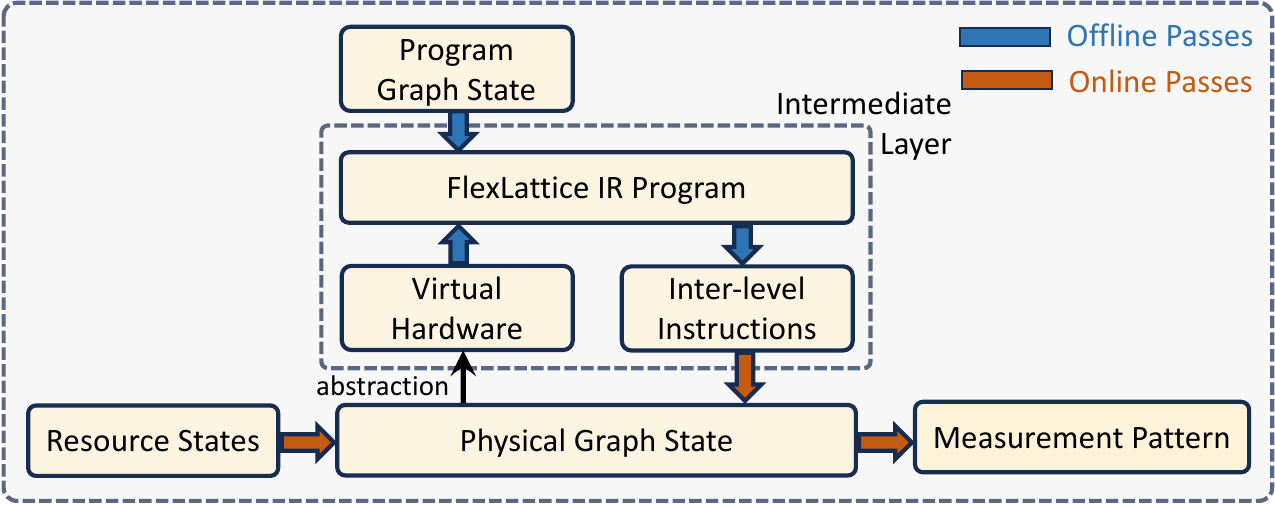}

    \caption{High-level design of OnePerc.}
    \label{fig:framework_design}
\end{figure}

To this end, we propose a randomness-aware compiler to efficiently scale up quantum computing on photonic systems, as illustrated in Fig.~\ref{fig:framework_design}. Our framework achieves concurrent real-time scalability and program execution efficiency through the combination of an online pass and an offline pass. The online pass handles real-time randomness in a scalable manner through percolation and reshaping. In particular, it provides a general fusion strategy for various resource states 
and exposes the reshaped physical graph states to programs by 
the abstraction of a virtual hardware. Motivated by the features of this virtual hardware, we propose a \emph{FlexLattice} intermediate representation (IR), which preserves the high-level program information and provides maximal optimization space supported by the virtual hardware. 
This allows offline passes to address the mismatch between program and physical graph states
by transforming the program to an efficient IR program, which can then be translated to intermediate-level instructions to guide real-time operations.

Our contributions in this paper are summarized as follows:
\begin{itemize}
    \item We propose a randomness-aware compiler for photonic quantum computing through a combination of online and offline passes, which are bridged by a novel FlexLattice IR facilitated with an intermediate-level instruction set.
    \item The online pass handles real-time randomness in a scalable manner by formation of long-range connectivity with various resource states and efficient reshaping of the random long-range connected structures. 
    
    \item The FlexLattice IR provides programs with an optimization space that balances the complexity of online structure reshaping and the flexibility of the reshaped structures. This enables an offline pass to enhance the efficiency of program execution through optimized mapping algorithms.
    
    \item Our evaluation demonstrates a significant outperformance over the efficient baseline in a scalable manner, implying a first-time concurrent achievement of scalability and efficiency in compilation of photonic quantum computing.

\end{itemize}

%% file: 03_background.tex
\section{Background}\label{sect:background}
\input{03_background_1}

\subsection{Photonic Platform} 
The weak interaction between photons, despite ensuring low cross-talk between photonic qubits, also poses significant challenges for realizing multi-qubit gates in the circuit model. Therefore, as a computing model that only requires measurements, MBQC~\cite{con-rev} emerges as highly suitable for photonic quantum computing. 
Besides the experimental demonstration of small-scale photonic MBQC \cite{mbqc-grover,mbqc-dj,mbqc-simon}, the photonic platform is rapidly scaling up with integrated waveguides and optical chips ~\cite{con-rev-217,con-rev-204, ferreira2022deterministic, con-rev-247, con-rev-252, interleaving}

Practical photonic hardware scales up by creating small  \emph{resource states}, e.g., 4-qubit, 6-qubit graph states, and connecting them through \emph{fusions}~\cite{fbqc}. In particular, identical resource states are periodically generated by an array of \emph{resource state generators} (RSGs) every cycle, with those generated in the same RSG cycle forming a 2D resource state layer (RSL).
Along with the time dimension, this creates a 3D array of resource states in the space-time. 

The resource states can then be merged probabilistically into larger graph states through (type II~\cite{linear_optics}) fusions, which can be regarded as concurrent measurements of $X\otimes Z$ and $Z\otimes X$, on two photonic qubits from different resource states. Resource states on the same RSL can fuse with their neighbors
by a spatial routing of photons, while 
resource states generated by the same RSG but on different RSLs
can fuse with each other by a temporal routing that controls the arrival times of photons at measurement devices.

 \begin{figure}[h]
    \centering
    \includegraphics[width=0.38\textwidth]{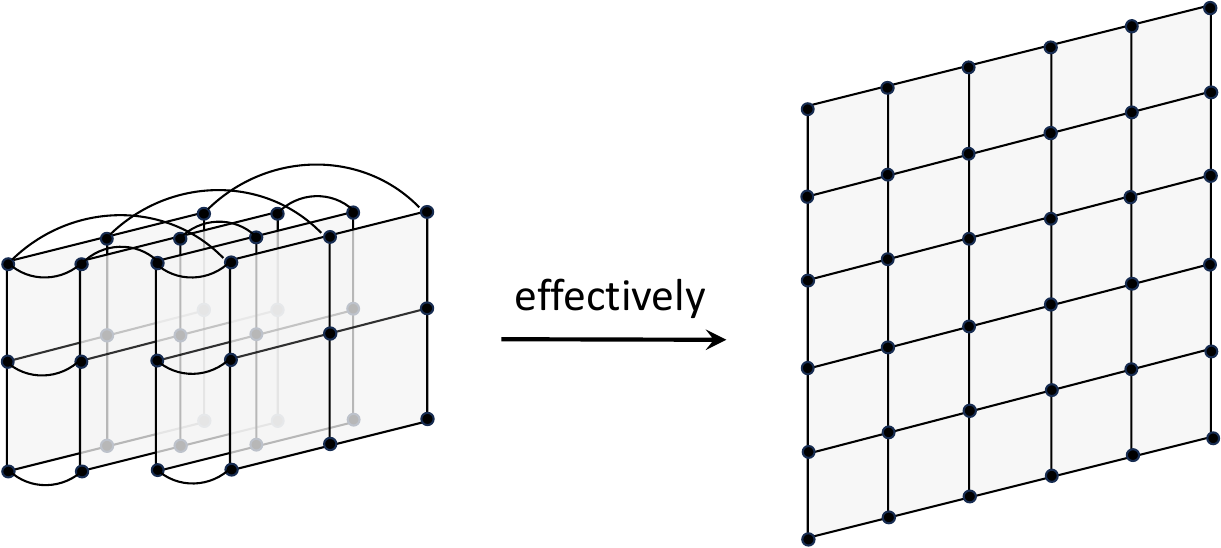}\\
    (a)\hspace{100pt}(b)
    \caption{Form a large RSL from multiple small RSLs.}
    \label{fig:3D_structure}
\end{figure}

With the advanced integrated silicon photonics, hardware components described above can operate on the scales characterized by GHz clock rates~\cite{GHz_1,GHz_2,GHz_3}, potentially leading to a time scale $\sim$ 1 ns for RSG cycles \cite{interleaving}. Spatial routing can be adjusted in every RSG cycle with switches, while temporal routing can be achieved by temporarily storing photonic qubits in a  high-capacity quantum memory known as \emph{delay lines}, realized by optical fiber technology. With a low transmission loss rate of $<5\%$ per km~\cite{fiber}, photons can have a lifetime of around 5000 RSG cycles in the delay lines. Moreover, the size of RSL is not completely constrained by the number of RSGs, but can be extended by leveraging the tradeoff between spatial and temporal fusions~\cite{interleaving}. For example, the large RSL depicted in Fig.~\ref{fig:3D_structure}(b) can be formed by fusing the edges of several small RSLs as depicted in Fig.~\ref{fig:3D_structure}(a), resembling folding a paper by twice (\ref{fig:3D_structure}(b)$\rightarrow$ \ref{fig:3D_structure}(a)).
With a photon lifetime around 5000 RSG cycles, this allows for an extension of RSL size by up to 5000 times.

However, as the key operation for merging resource states, fusions are intrinsically probabilistic. By allowing two ancilla photons, their success probability can be practically boosted to 75\%~\cite{0.78, 0.95with30}. While no conceptual limit has been found yet, so far the maximum known success probability attainable using linear optics is 78\% by injecting 8 ancilla single photons~\cite{0.78}. Reaching a higher success probability not only requires a larger number of ancilla photons but could also require the ancilla photons to be entangled. 
For example, it would take 30 entangled ancilla photons to reach a success probability over 95\%~\cite{0.95with30}.

%% file: 03_background_1.tex
\subsection{MBQC Basics}
\begin{figure}[h!]
    \centering
    \includegraphics[width=0.98\linewidth]{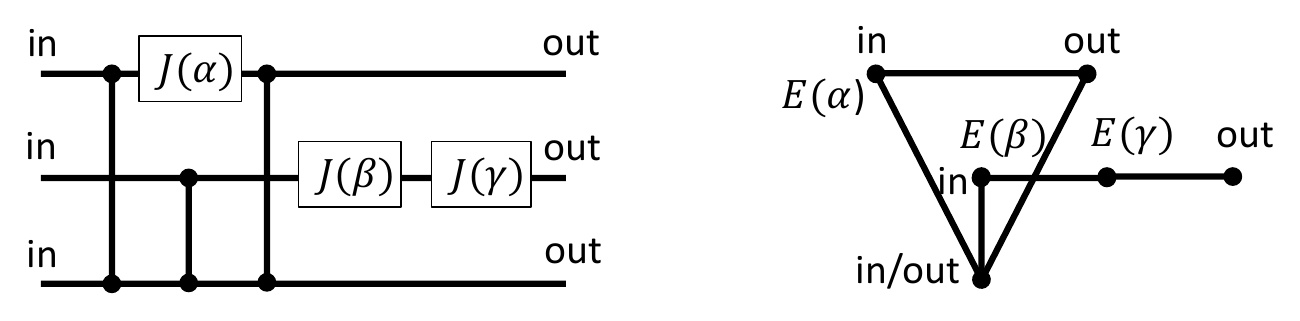}\\
    \hspace{0pt}(a)\hspace{100pt}(b)
    \caption{Translation from a circuit (a) to a measurement pattern on a graph state (b).}
    \label{fig:circuit_translation}
\end{figure}

MBQC is a universal but conceptually distinct computational model from the circuit model. In MBQC, computation is driven by 1-qubit projective measurements, rather than 1-qubit and 2-qubit gates, on an initial entangled state called \emph{graph state}, whose graph structure $G=(V,E)$ is determined by the quantum program~\cite{translation}. As exemplified in~\ref{fig:fusion_failures}(a), \ref{fig:circuit_translation}(b), \ref{fig:oneq+random}(a), each vertex in the graph state stands for a qubit, with the state 
formally defined as the eigenstates of operator\[s=X_i \bigotimes_{j\in n_i} Z_j, \quad\forall i\in V\] 
where $X_i, Z_j$ are the Pauli operators on qubit $i,j$ respectively, and $n_i$ is the set of neighboring qubits of $i\in V$ on graph $G$.
On the graph state, computation can be driven by a set of \emph{equatorial measurements} $E(\alpha)$, i.e., measurements on the X-Y plane of Bloch sphere at an angle $\alpha$, along with $Z$-measurements.
The measurement basis of each qubit is predetermined by the quantum program, known as a \emph{measurement pattern}, but are subject to a real-time adjustment according to the measurement outcomes of prior qubits, with the angles adjusted from $\alpha$ to $(-1)^s\alpha+t\pi$ where $s,t\in\{0,1\}$. This feed-forward mechanism is used to address the non-determinism of quantum measurement outcomes.

MBQC has the same computation power with the circuit model in the sense that they are both universal computing models. There is a straightforward translation~\cite{translation} from a circuit in the universal gate set $\{J(\alpha), \mathrm{CZ}\}$ into a measurement pattern on a graph state, 
where
\begin{align*}
J(\alpha)=
\begin{bmatrix}
1 & \mathrm{e}^{\mathrm{i} \alpha}\\
1 & -\mathrm{e}^{\mathrm{i} \alpha}
\end{bmatrix}  
\end{align*}
For example, Fig.~\ref{fig:circuit_translation} shows the translation from ~\ref{fig:circuit_translation}(a) to ~\ref{fig:circuit_translation}(b), each vertex in (b) representing a qubit, with `in' and `out' denoting their roles of being input or output qubits. It can be seen that the gates $J(\alpha)$, $J(\beta)$$, J(\gamma)$ are translated to equatorial measurements with corresponding angles, i.e., $E(\alpha)$, $E(\beta)$, $ E(\gamma)$, while the CZ gates are translated to edges of the graph states. This process can be rigorously described in ZX-calculus and optimized by available tools such as PyZX~\cite{kissinger2019pyzx}.

%% file: 04_motivation.tex
\section{Motivation and Overview}

In this section, we present a motivating example to demonstrate that a straightforward adaption of OneQ is insufficient to yield a scalable compiler in the presence of fusion failures. Then we provide an overview of our innovative randomness-aware compiler designed to effectively overcome these challenges.

\subsection{Motivating Example}
\begin{figure*}[t]
    \centering
    \includegraphics[width=0.96\textwidth]{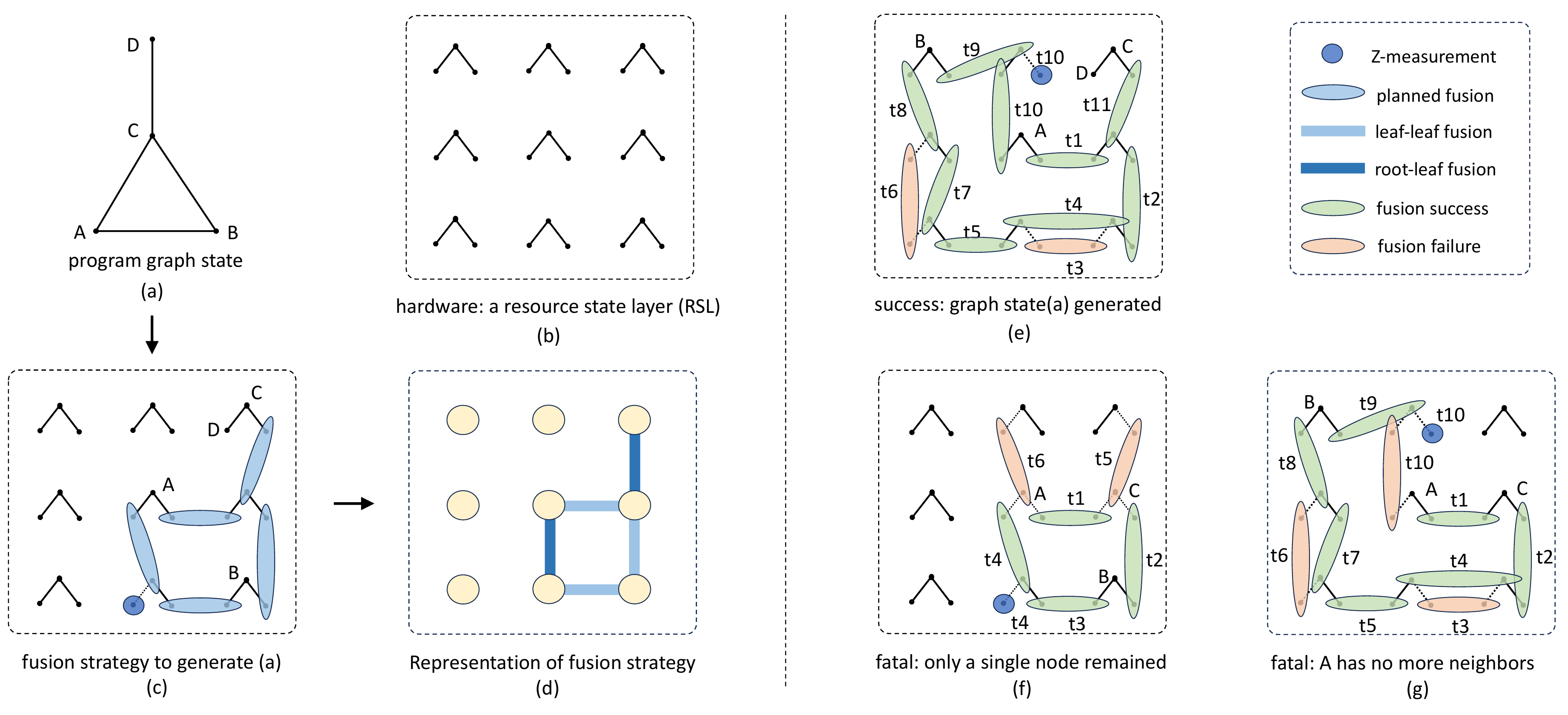}

    \caption{Why OneQ does not work.}
    \label{fig:oneq+random}
\end{figure*}

\paragraph{OneQ + Retry}
Fig.~\ref{fig:oneq+random}(c) illustrates OneQ's strategy with an example of generating a graph state in Fig.~\ref{fig:oneq+random}(a) from a single RSL depicted in Fig.~\ref{fig:oneq+random}(b), with the strategy represented more compactly by Fig.~\ref{fig:oneq+random}(d). Since the resource states have a star-like tree structure, we refer to the qubits of degree 1 as \emph{leaf qubits}, and those of degree $> 1$ 
as \emph{root qubits}.  
Light blue lines in Fig.~\ref{fig:oneq+random}(d) denote \emph{leaf-leaf fusions} (i.e., fusions between two leaf qubits) of the resource states (yellow circles), while dark blue lines denote \emph{root-leaf fusions} (i.e., fusions between a root and a leaf qubit).

To handle real-time randomness, a straightforward adaption is to introduce a retry mechanism. For example, the strategy in Fig.~\ref{fig:oneq+random}(c) can result in a dynamic implementation in Fig.~\ref{fig:oneq+random}(e) according to the fusion successes and failures (green and red ellipses), with these fusions performed sequentially from t1 to t11. If a fusion such as t3 fails, we retry the fusion using another two qubits at t4, and the same approach is applied to t6 and t7. This allows us to successfully generate the graph state in Fig.~\ref{fig:oneq+random}(a).

However, it is worth noting that some fatal failures may necessitate the retry of the entire compilation. For example, in Fig.~\ref{fig:oneq+random}(f), the triangular structure ABC is successfully generated from t1 to t4, but subsequent failures at t5 and t6 deplete the qubits in ABC, only leaving the isolated qubit B. In Fig.~\ref{fig:oneq+random}(g), a 5-qubit linear graph state forms from t1 to t9, which provides the potential for generating the triangle ABC if the fusion at t10 succeeds in fusing the two qubits at the line ends. Unfortunately, this fusion fails and consumes the last neighboring resource state of qubit A (except C), leaving A with no chance to fuse with other qubits. 

\paragraph{Critical Issues} From this example, we can find some critical issues of this dynamic retry mechanism. First, adapting to prior fusion outcomes necessitates a sequential execution of fusions. This considerably extends the processing time for each RSL, resulting in a time inefficiency as subsequent RSLs must wait for the completion of the current one. Second, since the decision-making process for responding to prior fusions occurs in real time, this extended processing time could exceed the limited lifetime of photons, especially for large RSLs. This would result in substantial photon loss, compromising the overall fidelity as computing scales up. Third, the frequent retries in real-time implementation lead to significant deviations from the planned strategy in Fig.~\ref{fig:oneq+random}(c). This undermines the benefits of the proactive planning, eroding the efficiency achieved by the mapping strategy of OneQ.

\subsection{Framework Overview}

Tolerating randomness in the compilation while maintaining efficiency presents a significant challenge. 
To address this, we propose an innovative framework that achieves scalability and efficiency simultaneously through a synergy of online and offline passes. The online pass prioritizes the real-time scalability by
maximizing the concurrency among fusions and the parallelism of the  associated path searching. The offline pass focuses on the efficient deployment of high-level program graph states onto the randomness-eliminated computing resource guaranteed by the online pass. The bridge between the online and offline passes is established through an intermediate software layer positioned between the low-level physical layer and the high-level program layer. This is achieved through a novel FlexLattice IR, along with an instruction set supported by the online pass and fulfilling the requirements of the offline pass.

\begin{figure}[h]
    \centering
    \includegraphics[width=0.44\textwidth]{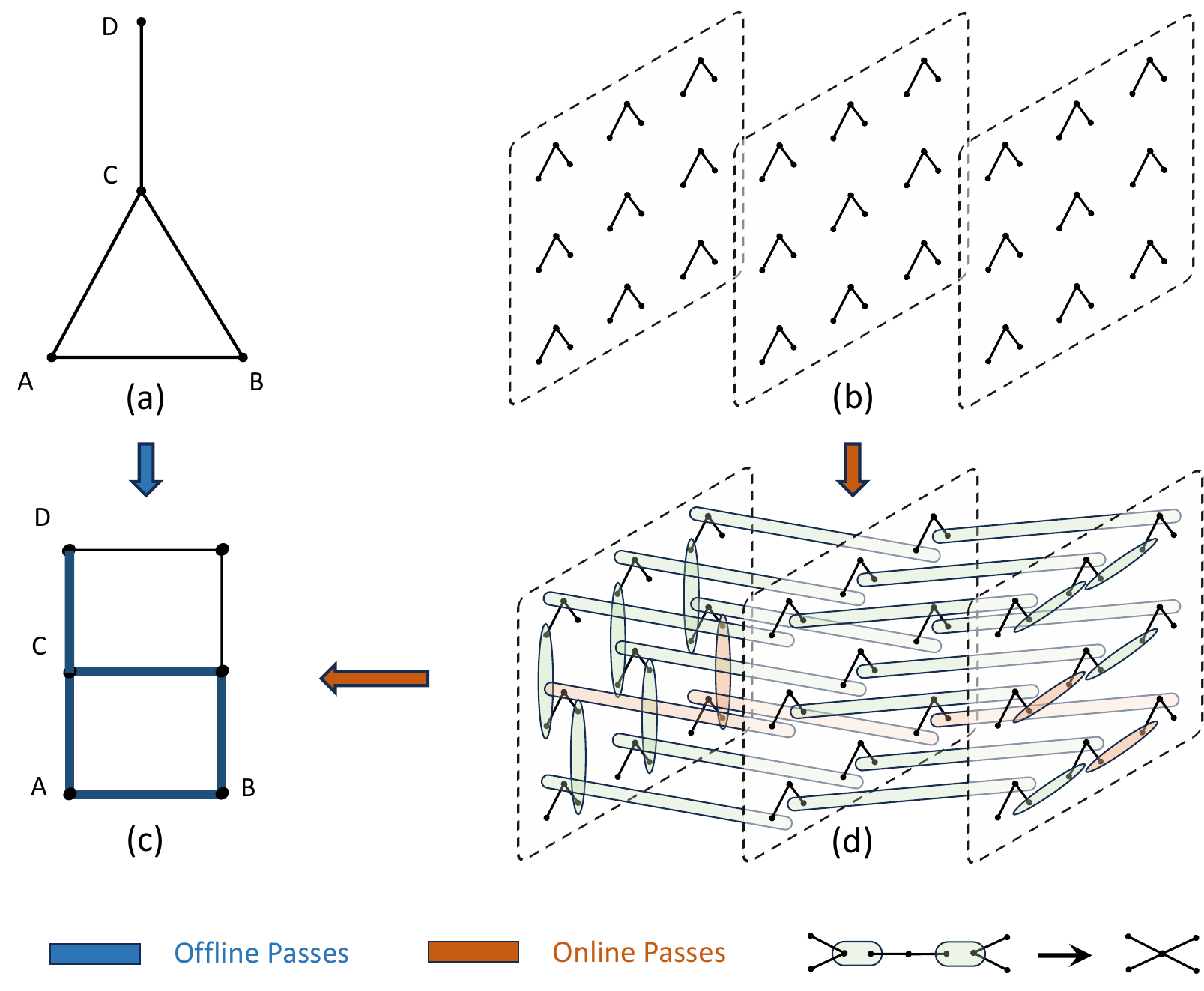}
    \caption{Overview of the compilation flow.}
    \label{fig:overview_example}
\end{figure}

To provide a concise overview, we exemplify the compilation flow by compiling a simple program graph state in Fig.~\ref{fig:overview_example}(a) onto the hardware in Fig.~\ref{fig:overview_example}(b), which is 3 layers of that in Fig.~\ref{fig:oneq+random}(b). Indeed, while Fig.~\ref{fig:oneq+random}(b) depicts only a single RSL, the incorporation of additional layers is both allowed and necessary for larger graph states. Steps (b)$\rightarrow$(d)$\rightarrow$(c) demonstrate the online pass, while step (a)$\rightarrow$(c) illustrates the offline pass. 
In the online pass, fusions are conducted concurrently in a predetermined pattern (Fig.~\ref{fig:overview_example}(d)) without individual retries of the failed ones, which eliminates the necessity for sequential operations. 
In this simple example, the resource states would result in a $3\times 3$ lattice if all fusions succeed, since the 3 resource states on the same locations of different layers would form a 4-degree star-like graph state (as depicted in the legend), while these star-like graph states would be joined into a lattice.
In the presence of fusion failures, the resulting physical graph state becomes a subgraph of the $3\times 3$ lattice,
which is then reshaped to a smaller lattice (Fig.~\ref{fig:oneq+random}(c)).
The target structure of the reshaping is program-agnostic, with its simple and regular structure facilitating the enhancement of real-time efficiency. When the fusion success probability exceeds the percolation threshold, this reshaping process attains near-deterministic success as the RSL size increases. This eliminates the necessity for repetitive retries of the entire compilation. With this near-determinism, the offline pass can be employed to improve the efficiency by mapping the program graph state compactly onto the reshaped lattice (bold blue lines in Fig.~\ref{fig:oneq+random}(c)).

Note that the compilation of general programs can be considerably more intricate than the example presented here. 
First, the fusion strategy among resource states is more complex than Fig.~\ref{fig:overview_example}(d). Specifically, it enables the formation of a 3D structure rather than 2D, being adaptable to various resource states and allowing collective retries with a small overhead. Second, the complexity of the reshaping algorithm is carefully reduced to enhance its real-time scalability. This is achieved by a modular design on each RSL that improves the parallelism of path searching. Third, the reshaping process is heterogeneous in the spatial and temporal dimensions, with the temporal dimension supporting connections both between adjacent layers and non-adjacent layers. These flexible connections provides a larger optimization space for the offline mapping than Fig.~\ref{fig:overview_example}(c). Forth, the online and offline passes are further bridged by posing a FlexLattice IR, which
guides the low-level operations by its translation to an instruction set.
For general programs, the compilation flow can be summarized as the following.

\begin{enumerate}
    \item Before program execution, an offline pass transforms the program graph state to an efficient FlexLattice IR, which is then translated to intermediate-level instructions to guide real-time operations (Section~\ref{sect:tech3}).
    \item During real-time execution, fusions between resource states are performed concurrently in a predetermined pattern, allowing collective retries of failed connections to improve the long-range-connectivity of the resulting physical graph state (Section~\ref{sect:tech1}).
    \item The resulting physical graph state is then reshaped to a 3D structure that fulfills the requirement of the IR program, with measurements performed on qubits according to the IR program (Section~\ref{sect:tech2}).
\end{enumerate}
The following sections (Section~\ref{sect:tech1}, \ref{sect:tech2}, \ref{sect:tech3}) will provide a bottom-up introduction to the framework.

%% file: 06_tech_1_revised.tex
\begin{figure*}[tp]
    \centering
    \includegraphics[width=0.96\textwidth]{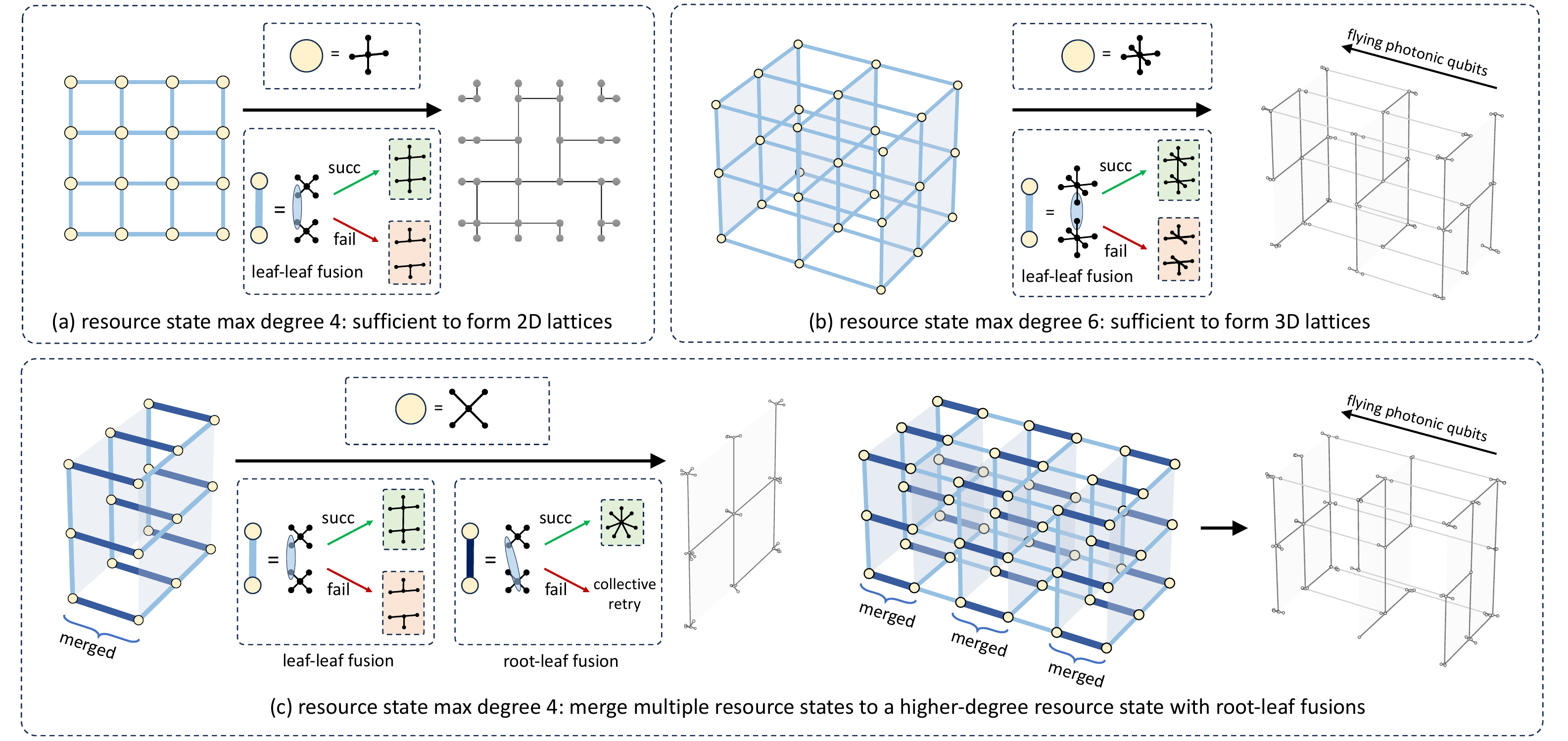}

    \caption{Fuse small resource states into 3D lattices.}
    \label{fig:dynamic_fusion}
\end{figure*}

\section{Resource State Fusion}
\label{sect:tech1}
In this section, we discuss the semi-static fusion strategy for generic star-like resource states. This strategy is static in that it is predetermined independent of high-level programs, yet semi-static in that it allows collective retries which induces only a constant overhead. In Fig.~\ref{fig:framework_design}, this corresponds to the online pass from resource states to physical graph states.

\subsection{Sufficient / Insufficient Degree} The predetermined strategy attempts to create a lattice structure from the resource states, which is straightforward when resource states have sufficient node degrees, i.e., the maximum degree in the resource states surpasses that in the lattice. For example, Fig.~\ref{fig:dynamic_fusion}(a) shows the strategy of forming a 2D square lattice using 4-degree resource states. 
On the left side, each light blue line represents a leaf-leaf fusion of the resource states (yellow circles). The right side displays the resulting physical graph state, with the consequences of successful and failed fusions depicted in the middle box. Similarly, Fig.~\ref{fig:dynamic_fusion}(b) demonstrates the case of forming a 3D cubic lattice from 6-degree resource states. 

As resource states on realistic hardware may lack sufficient degrees, such as forming 3D cubic lattices using 4-degree resource states, we can increase resource state degrees by merging multiple RSLs into one layer using root-leaf fusions, represented by the dark blue lines in Fig.~\ref{fig:dynamic_fusion}(c). Upon fusion success, while the two qubits in the fusion vanish, the two set of neighboring qubits of them would be connected in pairwise.
Hence a successful root-leaf fusion between two 4-degree resource states can generate a 7-degree graph state, which then has sufficient degree to form a 3D lattice.

\subsection{Removal of Irregular Structures}

\begin{figure}[h]
    \centering
    \includegraphics[width=0.22\textwidth]{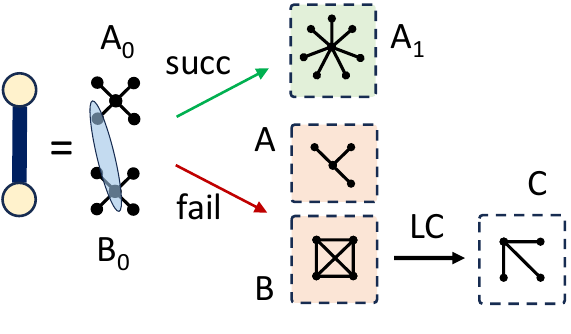}
    \caption{Root-leaf fusion failure.}
    \label{fig:root_leaf_failure}
\end{figure}

However, a failed root-leaf fusion may result in irregular cyclic structures in the generated graph state, leading to significant challenges for subsequent reshaping process. 
For example, a failed root-leaf fusion between two resource states $A_0$ and $B_0$ in Fig.~\ref{fig:root_leaf_failure} generates a star-like graph state $A$ and a fully connected cyclic graph state $B$. This is because a failed fusion on a qubit $v$ can be regarded as removing the qubit after a process of \emph{local complementation} (LC) on $v$, denoted as $\tau_v(G)$. Specifically, LC is defined as: among all neighbors of qubit $v$ in its resource state, if there was an edge between a pair of neighbors, then that edge is deleted; otherwise, an edge is added between that pair of neighbors.

To remove these cyclic structures, resource states with failed root-leaf fusions 
can be transformed to their local complementations of star-like structures (from $B$ to $C$ in Fig.~\ref{fig:root_leaf_failure}) by applying the following sequence of 1-qubit operators \cite{LCRule}
\[
U_v(G) = \exp\large(-\mathrm{i}\frac{\pi}{4}X_v\large)\prod_{u\in N_v} \exp\large(\mathrm{i}\frac{\pi}{4}Z_u\large)
\]
with $N_v$ being the neighbors of $v$ in $G$. Computation on these local complementation states is equivalent to that on the original states.
This is because we can interchange the orders of measurements and LC operators by adjusting the measurement bases, with the rules summarized in Theorem.~\ref{theorem_3}. Similarly, the LC operators can also be interchanged with fusion operations, with the rules summarized in Theorem~\ref{theorem_4}. Consequently, all LC operators can be postponed to the end of the computing process, 
which eliminates the necessity to implement them in the real-time.

\begin{theorem}
The local operator $U_{Z}^{\pm}=\exp(\pm \mathrm{i}\frac{\pi}{4} Z)$ or  $U_{X}^{\pm} = \exp(\pm \mathrm{i}\frac{\pi}{4} X)$ can be propagated through a $Z-$measurement or a 1-qubit equatorial measurement on the Bloch sphere, i.e., a measurement in the basis of $\cos\phi X + \sin\phi Y$ where $\phi\in[0,2\pi)$, by a change of measurement basis. 

\begin{proof}

When measuring a 1-qubit state $\ket{\psi}$ along $A$-basis, the state collapses to $\ket{\psi'}\equiv\mathrm{M}_{[A]}\ket{\psi}=\frac{\mathbb{I}\pm A}{2}\ket{\psi}$, with the sign $\pm$ determined by the measurement outcome, 0 or 1. Therefore,
    \begin{align*}
        \mathrm{M}_ Z U_Z^\pm &=U_Z^\pm\mathrm{M}_Z\\
        \mathrm{M}_ Z U_X^\pm&=U_X^\pm\mathrm{M}_{[\mp Y]}\\
        \mathrm{M}_ {[\cos\phi X + \sin\phi Y]}U_Z^\pm
        &= U_Z^\pm\mathrm{M}_{[\pm(\cos\phi Y - \sin\phi X)]}\\
        \mathrm{M}_ {[\cos\phi X + \sin\phi Y]}U_X^\pm
        &= U_X^\pm \mathrm{M}_{[\cos\phi X \pm \sin\phi Z]}\qedhere
    \end{align*}
\end{proof}
\label{theorem_3}
\end{theorem}

\begin{theorem}

The local operator $U_Z^\pm=\exp(\pm \mathrm{i}\frac{\pi}{4} Z)$ or $U_X^\pm=\exp(\pm \mathrm{i}\frac{\pi}{4} X)$ can be propagated through a 2-qubit $XZ,ZX$ fusion, i.e., a joint measurement of $X_1Z_2,Z_1X_2$ on qubit 1 and qubit 2, by a change of fusion basis. 

\begin{proof}
When measuring a 2-qubit state $\ket{\psi}$ along basis $A_1 B_2$, the state collapses to $\ket{\psi'}\equiv\mathrm{M}_{[A_1 B_2]}\ket{\psi}=\frac{\mathbb{I}\pm A_1 B_2}{2}\ket{\psi}$, with $\pm$ determined by the measurement outcome, 0 or 1. Therefore
\begin{align*}
\mathrm{M}_{[X_1 Z_2]}&\mathrm{M}_{[Z_1 X_2]} U_{Z_1}^{\pm_1}U_{Z_2}^{\pm_2}=U_{Z_1}^{\pm_1}U_{Z_2}^{\pm_2} \mathrm{M}_{[\pm_1 Y_1 Z_2]}\mathrm{M}_{[\pm_2 Z_1 Y_2]}\\
\mathrm{M}_{[X_1 Z_2]} &\mathrm{M}_{[Z_1 X_2]} U_{X_1}^{\pm_1}U_{X_2}^{\pm_2}=U_{X_1}^{\pm_1}U_{X_2}^{\pm_2} \mathrm{M}_{[\mp_1 Y_1 X_2]}\mathrm{M}_{[\mp_2 X_1 Y_2]}\\
\mathrm{M}_{[X_1 Z_2]} &\mathrm{M}_{[Z_1 X_2]} U_{Z_1}^{\pm_1}U_{X_2}^{\pm_2}=U_{Z_1}^{\pm_1}U_{X_2}^{\pm_2} \mathrm{M}_{[\pm_1 \mp_2 Y_1 Y_2]}\mathrm{M}_{[Z_1 X_2]}\qedhere
\end{align*}
\end{proof}
\label{theorem_4}
\end{theorem}

\subsection{Collective Feed-forward}
The semi-static fusion strategy allows collective feed-forward in the granularity of RSL, which can be pipelined to reduce the overhead. On one hand, the propagation of LC operators through measurements and fusions requires an adaptive adjustment of measurement and fusion bases. With this dependency, each RSL are fused in two batches: a batch of root-leaf fusions and a batch of leaf-leaf fusions. On the other hand, the connectivity of the physical graph state can be enhanced by retries of the failed connections, including retrying failed leaf-leaf fusions with redundant degrees (e.g., with the $7^\mathrm{th}$ degree of $A_1$ in Fig.~\ref{fig:root_leaf_failure}) and retrying failed root-leaf fusions with remaining degrees (e.g., with $A$ and $C$ in Fig.~\ref{fig:root_leaf_failure}).
In this way, each RSL may undergo more batches of fusions. However, since earlier batches of later RSLs can be conducted concurrently with later batches of earlier RSLs,
this only introduces a constant overhead to program execution.

%% file: 06_tech_2_revised.tex
\section{Random State Reshaping}
\label{sect:tech2}
\begin{figure*}[tp]
    
    \centering
    \includegraphics[width=0.96\textwidth]{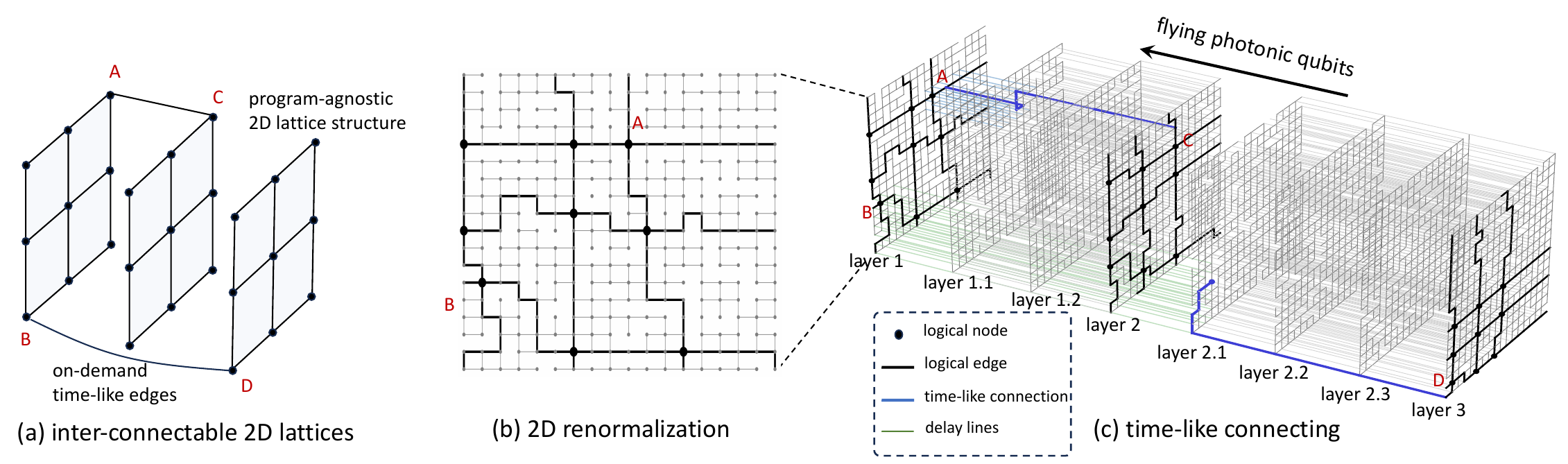}
    \caption{(2+1)-D reshaping for handling random graph states generated by fusions.}
    \label{fig:(2+1)-d renorm}
\end{figure*}

In this section, we delve into the reshaping of physical graph states, which is characterized by a (2+1)-D design, motivated by the continuous generation of RSLs over time and the presence of delay lines.
In Fig.~\ref{fig:framework_design}, this corresponds to the online pass from physical graph states to measurement patterns.

\subsection{Efficient 2D Renormalization} On each (merged) RSL, we apply a process known as renormalization \cite{2D_renorm}, which reshapes the largest connected component of the physical graph state to a coarse-grained 2D lattice.
The key to its viability lies in the percolation phenomenon~\cite{3GHZ,perc_threshold,2D_renorm}. 
That is, when the fusion success probability exceeds a certain threshold, the random physical graph state undergoes a phase transition from short-range connectivity to long-range connectivity, leading to the largest connected component reaching a comparable size with the original graph state. 
Since fusions on each RSL are constrained as a squared lattice, the percolation threshold is only 0.5~\cite{square_lattice_perc}, lower than the achievable fusion success probability.

Identifying intersections of horizontal and vertical paths in the largest connected component reveals a coarse-grained square lattice, represented by bold nodes and edges in Fig.~\ref{fig:(2+1)-d renorm}(b). 
This is achieved in the following way. We search for vertical paths from left to right and horizontal paths from bottom to top, enforcing distinct vertical or horizontal paths to maintain a separation of at least one qubit. When searching for vertical (horizontal) paths, a connectivity check is conducted between nodes at the top (left) and bottom (right), facilitated by a disjoint-set data structure to reduce the complexity. Upon confirming connectivity, a breadth-first search (BFS) is applied to determine the shortest path, ensuring it remains free of self-tangling. To further prevent tangling between vertical and horizontal paths, we remove the surrounding qubits of each identified path after discovery, preventing their interference with subsequent searches. Considering the removals, an alternating search of vertical and horizontal paths emerges
as an effective searching order.

\begin{figure}[h]
    \centering
    \includegraphics[width=0.4\textwidth]{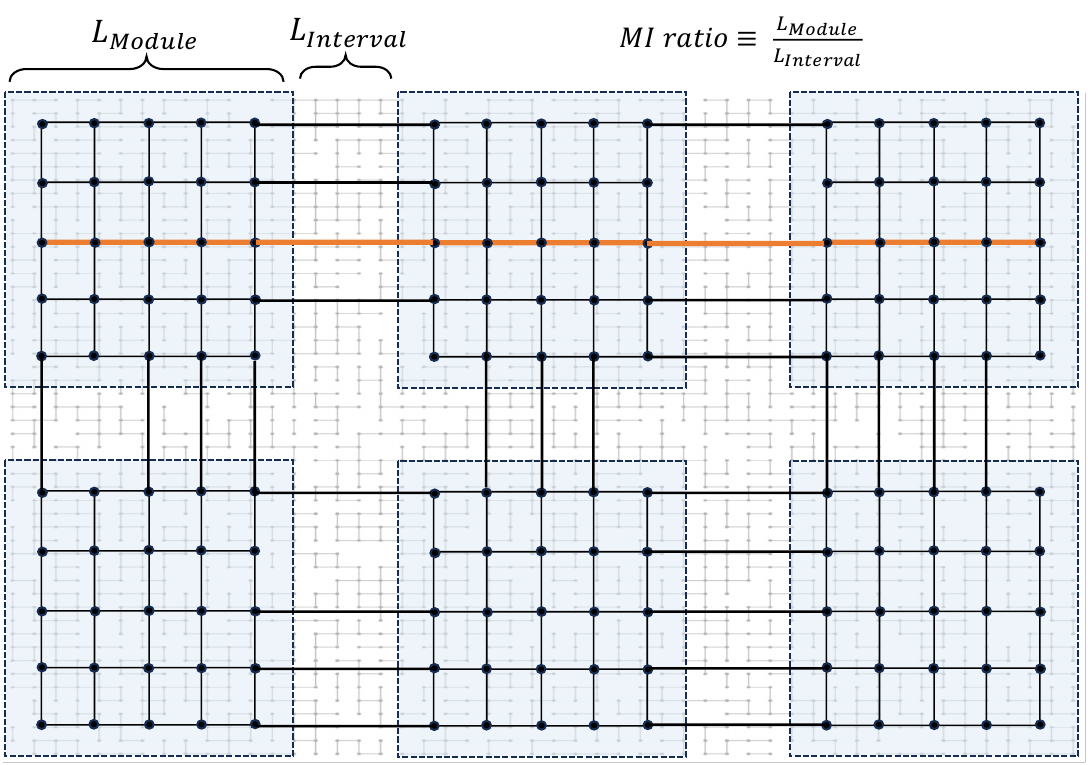}
    \caption{Modular renormalization.}
    \label{fig:modular_renormalization}
\end{figure}

To improve real-time scalability, the 2D renormalization is designed to allow \textbf{modularity}, with areas on the RSL renormalized concurrently and then joined together. As shown in Fig.~\ref{fig:modular_renormalization}, the RSL is divided into several modules of size $L_{Module}\times L_{Module}$, with some intervals of length $L_{interval}$ left in between for joining the modules by connected paths. With the path searching algorithm above, the complexity of a modular 2D renormalization is $O(L_\mathrm{module}^2)\sim O(N^2/m)$, where $m$ is the number of modules. Since an entire path can only be established if all inter-module paths involved are successful (e.g., the orange path), the potential for failed inter-module paths could lead to the renormalized lattice size being smaller than the total size of all individual modules.
A suitable ratio of $L_{Module}$ and $L_{Interval}$, defined as \emph{MI ratio}, can help mitigate this resource overhead.

\subsection{Flexible Time-like Connections} 
Nodes on the renormalized 2D lattices can be connected along the time dimension, referred to as \emph{time-like connections}. Connections between adjacent 2D lattices and across non-adjacent 2D lattices are called \emph{adjacent-layer} connections and \emph{cross-layer} connections, respectively. Before program execution, the connections to establish are given by the IR program as a 3D graph, as illustrated in Fig.~\ref{fig:(2+1)-d renorm}(a). 

The process of generating this 3D graph is illustrated in Fig.~\ref{fig:(2+1)-d renorm}(c) on 8 RSLs, with the adjacent-layer connection $AC$ and the cross-layer connection $BD$ implemented through the bold blue paths. 
This involves an attempt of 2D renormalization on each RSL. The successful ones then serve as \emph{logical layers}, indexed by integers in Fig.~\ref{fig:(2+1)-d renorm}(c), with the renormalized nodes on them referred to as \emph{logical nodes}.  In contrast, the RSLs with failed renormalization serve as \emph{routing layers}, which are indexed by decimals in Fig.~\ref{fig:(2+1)-d renorm}(c). The renormalization on an RSL is considered successful if:

\begin{enumerate}
    \item The renormalized 2D lattice reaches a target size, which is equivalent to a choice of average node size, where 
    \[\mathrm{average\_node\_size} \equiv \frac{\mathrm{RSL\_size} }{\mathrm{renormalized\_lattice\_size}}\]

    \item The RSL can establish all necessary time-like connections with prior logical layers through the following procedure.
\end{enumerate}

To establish a time-like connection between two nodes, a set of physical qubits around the preceding node are fused with corresponding qubits on a correct subsequent RSL. 
For adjacent-layer connections such as $AC$, the qubits around $A$ are directly fused to the next RSL, which is layer 1.1 in Fig.~\ref{fig:(2+1)-d renorm}(c). For cross-layer connections such as $BD$, the qubits around $B$ are temporarily stored in delay lines, depicted by the green thin lines in Fig.~\ref{fig:(2+1)-d renorm}(c), until they can be fused to layer 2.1, which is the first RSL between the current attempting RSL (layer 3) and its prior logical layer (layer 2). 
Subsequently, a path searching between the two nodes is conducted within the physical graph state, exemplified by the bold blue lines $AC$ and $BD$ in Fig.~\ref{fig:(2+1)-d renorm}(c). Again, this is achieved by a connectivity check utilizing a disjoint-set data structure and a BFS for the shortest path.
If the connectivity check yields a negative result, it indicates that the current RSL fails to meet the second condition and would become a routing layer.
It is worth mentioning that the reshaping process can tolerate photon loss, since a fusion is considered as successful only if both two photons are detected. 
Effectively, the presence of photon loss causes a reduction of the fusion success probability, possibly leading to more routing layers between logical layers. 

In contrast to the logical layers, all qubits of each routing layer are directly fused with their next RSL, as depicted by the grey thin lines in Fig.~\ref{fig:(2+1)-d renorm}(c). This is because before obtaining the next successful renormalization, we can’t predict where the logical node would locate and which fusions around it would succeed. Moreover, in contrast to the simple case in Fig.~\ref{fig:(2+1)-d renorm} where there is only one connection between layer 1 and layer 2, in practice we may need to establish multiple connections between logical layers. This makes it even harder to predict which fusions are redundant before executing the fusions. 

%% file: 06_tech_3_revised.tex
\section{Offline Optimization with IR}
\label{sect:tech3}
In this section, we introduce the virtual hardware, FlexLattice IR, offline mapping and intermediate-level instructions. This covers the offline pass in the compilation flow (Fig.~\ref{fig:framework_design}).

Before program execution, the mismatch between program graph states and physical graph states can be addressed by mapping the program graph state onto the virtual hardware, leading to an IR graph state that maintains the high-level program information. This IR program can then be transformed to a set of intermediate-level instructions, which guides real-time physical operations through the reshaping algorithm above.

\subsection{Virtual Hardware}
\begin{figure}[h]
    \centering
    \includegraphics[width=0.43\textwidth]{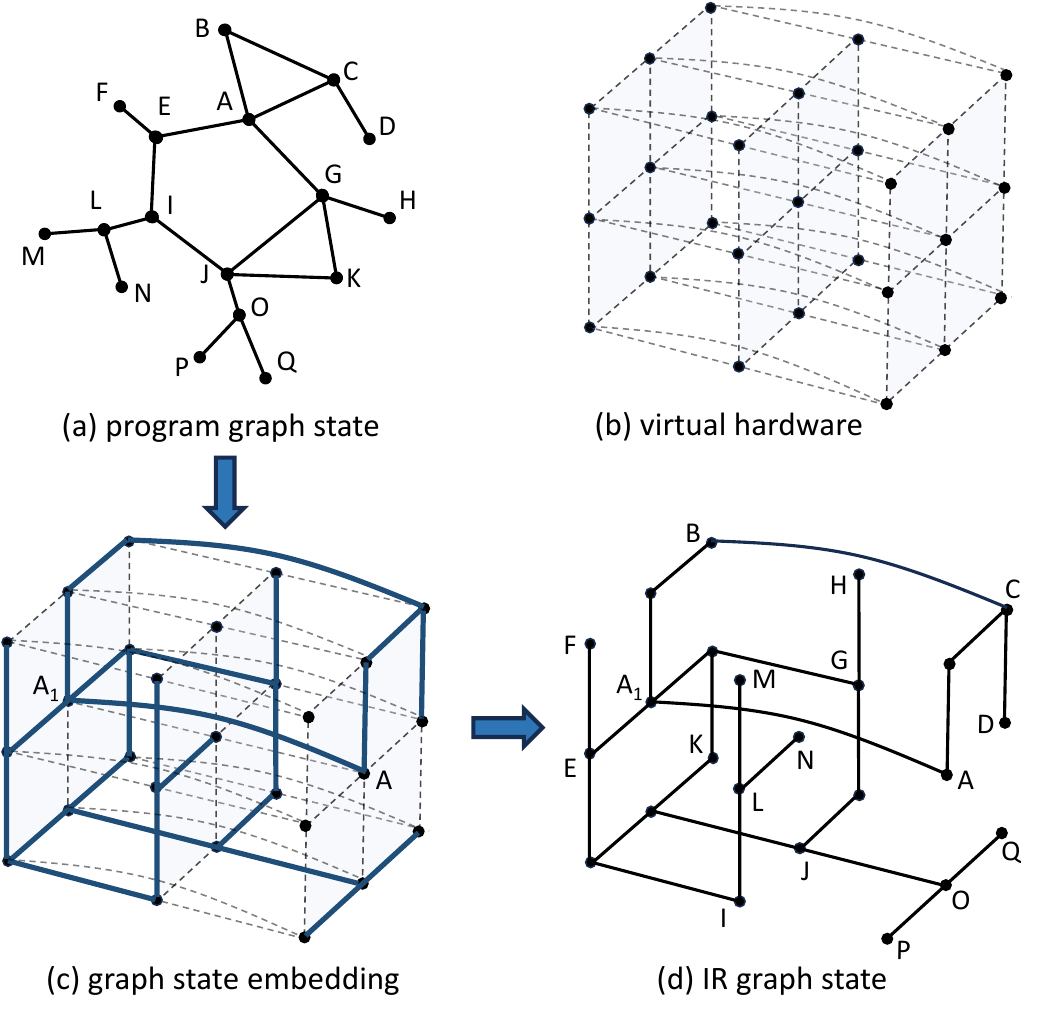}
    \caption{Offline mapping onto the virtual hardware.}
    \label{fig:virtual_hardware}
\end{figure}

The virtual hardware abstracts the adjustable structures supported by the reshaping algorithm. It pocesses a (2+1)-D structure,
characterized by the following features, as illustrated in Fig.~\ref{fig:virtual_hardware}(b).

\begin{enumerate}
    \item The virtual hardware consists of consecutive layers of 2D lattices in a fixed size, with a virtual memory located on each 2D coordinate.
    \item Nodes on the same 2D coordinate of different layers, either on adjacent or non-adjacent layers, can be connected along the third dimension, with the connections between non-adjacent layers realized by temporary storage of nodes in the virtual memory.
    \item Each connection within or between 2D layers can be enabled or disabled on demand, but each node can have at most one connection with preceding layers and at most one connection with subsequent layers.

\end{enumerate}

While this virtual hardware can be used to generate 3D cluster states (i.e., lattice-like graph states), which serve as the universal computing resource of MBQC in previous work~\cite{mbqc2009}, it is more advantageous in the following aspects. First, an individual connection can be flexibly enabled or disabled without removing any logical node or affecting other edges. This is in contrast to the cluster state, wherein the removal of edges is usually achieved by removing involved vertices and all their edges.
Second, the connections among 2D layers exhibit greater flexibility than cluster states. Specifically, inter-layer connections between nodes on the same 2D coordinates extend beyond adjacent layers, encompassing cross-layer connections as well.

\subsection{FlexLattice IR and Offline Mapping}
With this virtual hardware, graph state mapping algorithms such as that in OneQ can be utilized as an offline pass to enhance the efficiency of program execution. Specifically, the mapping onto virtual hardware transforms a program graph state to an equivalent IR program with compatible structure with the virtual hardware, which is referred to as a \emph{FlexLattice} IR based on its structural features. This process is illustrated by Fig.~\ref{fig:virtual_hardware}(a) $\rightarrow$ (c) $\rightarrow$ (d).

To further improve the mapping efficiency and scale to larger programs, we extend OneQ's mapping algorithm with three optimizations.
\begin{itemize}
    \item First, to map graph nodes as early as possible, we replace the static partition in OneQ with a dynamic scheduling. Specifically, we analyze the dependency among graph state qubits~\cite{measurement_calculus}, representing it with a directed acyclic graph (DAG) and updating the front layer of the DAG as nodes are consumed by the mapping.

    \item Second, to reserve enough space for routing and avoid node congestion, we enforce an upper-limit for the occupancy of incomplete nodes on each virtual hardware layer (25\% by default), with incomplete nodes defined as those mapped nodes whose edges are not all mapped yet. 

    \item Third, to mitigate the increasing demand on classical memory for graph information storage, we propose a refresh mechanism, which periodically retrieves all nodes stored in the virtual memory, refreshing them by mapping onto multiple layers of the virtual hardware, and then storing them again.

\end{itemize}

\subsection{Instruction Set}
A FlexLattice IR program can be executed by transforming to a set of intermediate-level instructions, which guides the real-time physical operations to generate necessary connections among logical nodes through the reshaping algorithm in Section~\ref{sect:tech2}.  By default, qubits in the physical graph state are subject to Z-measurements, which means that edges are disabled on the virtual hardware unless explicitly enabled by the intermediate-level instructions.
We list the intermediate-level instructions as the following, with nodes in the high-level program graph state denoted as \texttt{g\_node} and nodes on the virtual hardware denoted as \texttt{v\_node}.

\vspace{-5pt}
\begin{align*}
&\texttt{map\_v\_node(v\_node, g\_node)}\\
&\texttt{make\_v\_node\_ancilla(v\_node)}\\
&\texttt{store\_v\_node(v\_node)}\\
&\texttt{retrieve\_v\_node(v\_node, position)}\\
&\texttt{enable\_spatial\_v\_edge(v\_node, adjacent\_v\_node)}\\
&\texttt{enable\_temporal\_v\_edge(v\_node, adjacent\_v\_node)}
\end{align*}

By \texttt{map\_v\_node()} and \texttt{make\_v\_node\_ancilla()}, a virtual node can be mapped by a \texttt{g\_node} or used as an ancilla node to facilitate routing. In the former case, the physical qubit corresponding to \texttt{v\_node} will be measured in the basis of the \texttt{g\_node}, while in the latter case, it will be measured in $X$- or $Y$-basis to play as a wire (depending on whether the wire length is even or odd). By \texttt{store\_v\_node()} and \texttt{retrieve\_v\_node()}, a virtual node can also be stored into or retrieved from the virtual memory by pushing or poping its surrounding physical qubits to or from the delay lines. By \texttt{enable\_spatial\_v\_edge()}, a spatial edge between adjacent nodes on the same layer can be enabled by setting associated qubits to $X$- or $Y$-measurements. 

By \texttt{enable\_temporal\_v\_edge()}, a temporal edge between logical nodes at the same coordinate of adjacent layers can be enabled.
Establishment of a cross-layer edge between  layer $m$ and layer $n\enspace(>m)$ can be realized through the combination of three instructions: storing the node at layer $m$ into the virtual memory, retrieving it at layer $n-1$, and enabling a temporal edge between layer $n-1$ and layer $n$. For example, the cross-layer temporal edge between ancilla node $A_1$ at (1,1,0) and graph node $A$ at (1,1,2) in Fig.~\ref{fig:virtual_hardware}(d) can be implemented with the instructions below. Note that retrieving \texttt{v\_node} at layer $n-1$ does not conflict with the original \texttt{v\_node} at layer $n-1$ (i.e., node $N$ in Fig.~\ref{fig:virtual_hardware}(d)). This is because the original node at layer $n-1$ would not have an edge with layer $n$, since each node in a FlexLattice IR has at most one edge with preceding layers. This implies that the original node will either have no further edges or will be stored in the virtual memory at layer $n-1$.

\vspace{-5pt}
\begin{align*}
&\texttt{make\_v\_node\_ancilla((1, 1, 0))}\\
&\texttt{store\_v\_node((1, 1, 0))}\\
&\texttt{...}\\
&\texttt{retrieve\_v\_node((1, 1, 0), (1, 1, 1))}\\
&\texttt{enable\_temporal\_v\_edge((1, 1, 1), (1, 1, 2))}\\
&\texttt{map\_v\_node((1, 1, 2), A)}
\end{align*}

%% file: 07_evaluation.tex
\section{Evaluation}\label{sect:eval}

\subsection{Experiment Setup}\label{sect:expset}

\paragraph{Baseline} We compare the performance of our framework with the efficient photonic MBQC compiler OneQ. 
Since OneQ is not able to handle fusion failures, we employ it with a repeat-until-success strategy.
Specifically, for each RSL we conduct the fusions instructed by OneQ repeatedly until all fusions are successful. Subsequently, the successful RSL is fused with its preceding RSLs. If failures occur in the inter-RSL fusions, the entire compilation is restarted and repeated until success. 

\begin{table}[h]
    \centering
      \caption{Benchmark Programs.}
      
    \resizebox{0.48\textwidth}{!}{
        \renewcommand*{\arraystretch}{1}
        \begin{footnotesize}
        \begin{tabular}{|p{1.7cm}|p{1.5cm}|p{1.6cm}|p{1.6cm}|}  \hline
        Fusion Success Rate & \#Qubits & Virtual Hardware Size & RSL Size \\ \hline
        \multirow{3}{*}{0.90}  
        & 4 & 2x2 & 24x24\\ 
        \cline{2-4}  & 9 & 3x3 & 36x36\\

        \cline{2-4} & 25 & 5x5 & 60x60 \\
        \hline
         \multirow{4}{*}{0.75}  & 4 & 2x2 & 48x48\\
        \cline{2-4}  & 25 & 5x5 & 120x120 \\
        \cline{2-4} & 64 & 8x8 & 192x192 \\
        \cline{2-4} & 100 & 10x10 & 240x240 \\
        \hline

        \end{tabular}
        \end{footnotesize}
    }
    \label{tab:benchmark}
\end{table}

\paragraph{Metrics} 
Aligning with OneQ, we evaluate the performance of compilation with two metrics: the number of consumed RSL, denoted by \#RSL, and the number of required fusions, denoted by \#fusion.
In particular, a smaller \#RSL indicates less execution time of the program and less chance for photon loss, while a smaller \#fusion implies less operations and less chance for error occurrence.

\paragraph{Photonic Hardware Model} We adopt the same photonic hardware architecture with OneQ, as introduced in Section~\ref{sect:background}. In the main experiment (Table~\ref{tab:evaluation}, \ref{tab:refresh}), the comparison with OneQ is performed on 4-qubit star-like resource states, with the sizes of hardware for different benchmarks listed in Table~\ref{tab:benchmark}. Experiments for further analysis are conducted with 7-qubit star-like resource states, which naturally have sufficient degrees for forming 3D lattice-like graph states.

\paragraph{Benchmark Programs} 
We select a set of benckmark programs including Quantum Approximate Optimization Algorithm (QAOA), Quantum Fourier transform (QFT), Ripple-Carry Adder (RCA)~\cite{RCA} and Variational Quantum Eigensolver (VQE). 
For QAOA,
we choose the graph maxcut problem on randomly generated graphs. Specifically, the graphs are generated by randomly connecting half of all its possible edges.
For VQE, we follow the commonly used full-entanglement ansatz, which proves to be an expressive ansatz \cite{qiskit_vqe, vqe_ansatz}.
In table~\ref{tab:benchmark}, we list the benchmarks 
with their numbers of qubits in the circuit representation. 
We also list the sizes of virtual hardware layers, which are chosen to correspond with the qubit quantities, along with the required sizes of RSLs needed to generate them, which are determined through Fig.~\ref{fig:scalability}, as explained later.

\begin{table*}[tp]
    \centering
    \caption{The results of OnePerc and its relative performance to the baseline.}
    \resizebox{0.96\textwidth}{!}{
        \renewcommand*{\arraystretch}{1.06}
        \begin{footnotesize}
            \begin{tabular}{|p{1.8cm}|p{1.8cm}|p{1.8cm}|p{1.8cm}|p{1.8cm}|p{1.8cm}|p{1.9cm}|p{1.8cm}|}
        
        \hline
        Fusion Success Rate& Benchmark Name & OneQ \#RSL & OnePerc \#RSL & \#RSL Improv.  & OneQ \#Fusion & OnePerc \#Fusion & \#Fusion Improv.  \\
        \hline
         \multicolumn{1}{|c|}{\multirow{12}{*}{\parbox{2cm}{0.90\\(hyper-advanced)}}}  
         & QAOA-4 & 304 & 84 & 3.62 & 13,990 & 117,664 & 0.12\\
        \cline{2-8} & QFT-4 & 3,759 & 174 & 21.59 & 180,634 & 274,155 & 0.66\\
        \cline{2-8} & RCA-4 & 3,107 & 237 & 13.11 & 63,814 & 373,646 & 0.17\\
        \cline{2-8} & VQE-4 & 56 & 22 & 2.55 & 1,707 & 33,526 & 0.05 \\ 
        \cline{2-8} & QAOA-9 & \multirow{8}{*}{$>10^6$} & 240 & $>10^3$ & \multirow{8}{*}{$>10^{10}$} & 855,354 & $>10^4$\\
        \cline{2-2}\cline{4-5}\cline{7-8} & QFT-9 &  & 570 & $>10^3$ &  & 2,031,813 & $>10^3$\\
        
        \cline{2-2}\cline{4-5}\cline{7-8} & RCA-9 &  & 1,017 & $>10^2$ &  & 3,627,950 & $>10^3$\\
        
        \cline{2-2}\cline{4-5}\cline{7-8} & VQE-9 &   & 156 & $>10^3$ &  & 555,065 & $>10^4$\\
        
        \cline{2-2}\cline{4-5}\cline{7-8} & QAOA-25 &  & 768 & $>10^3$ &  & 7,637,711 & $>10^3$\\ 
        \cline{2-2}\cline{4-5}\cline{7-8} & QFT-25 &  & 2,418 & $>10^2$ &  & 24,065,102 & $>10^2$\\ 
        
        \cline{2-2}\cline{4-5}\cline{7-8} & RCA-25 &  & 3,111 & $>10^2$ &  & 30,962,172 & $>10^2$\\
        \cline{2-2}\cline{4-5}\cline{7-8} & VQE-25 &  & 705 & $>10^3$ &  & 7,010,656 & $>10^3$\\

        \hline
         \multicolumn{1}{|c|}
         {\multirow{12}{*}{\parbox{2cm}{0.75\\(practical)}}}  & QAOA-4 & 1,708 & 48 & 35.58 & 119,731 & 169,431 & 0.71\\
        \cline{2-8} & QFT-4 & $>10^6$ & 210 & $>10^3$ & $>10^{10}$ & 746,977 & $>10^4$\\
        \cline{2-8} & RCA-4 & $>10^6$ & 201 & $>10^3$ & $>10^{10}$ & 714,835 & $>10^4$\\
        \cline{2-8} & VQE-4 & 1,017 & 23 & 44.22 & 25,354 & 96,332 & 0.26\\

        \cline{2-8} & QAOA-25 & \multirow{8}{*}{$>10^6$} & 882 & $>10^3$ & \multirow{8}{*}{$>10^{10}$} & 19,743,350 & \multirow{8}{*}{$>10$}\\

        \cline{2-2}\cline{4-5}\cline{7-7} & QFT-25 &  & 2,271 & $>10^2$ &  & 50,835,771 & \\
        \cline{2-2}\cline{4-5}\cline{7-7} & RCA-25 & & 3,252 & $>10^2$ &  & 72,795,212 & \\
        \cline{2-2}\cline{4-5}\cline{7-7} & VQE-25 &  & 759 & $>10^3$ &  & 17,292,345 &  \\
        \cline{2-2}\cline{4-5}\cline{7-7} & QAOA-64 &  & 3,339 & \multirow{4}{*}{$>10^2$} &  & 191,341,276 & \\
        \cline{2-2}\cline{4-4}\cline{7-7} & QFT-64 &  & 9,000 &  &  &  515,801,985 & \\
        \cline{2-2}\cline{4-4}\cline{7-7} & RCA-64 &  & 9,324 &  &  & 534,311,489  & \\
        \cline{2-2}\cline{4-4}\cline{7-7} & VQE-64 &  & 3,042 &  &  & 174,321,702  & \\
        \hline

    \end{tabular}
    \end{footnotesize}
    }
    \label{tab:evaluation}
\end{table*}

\begin{table}[h!]
    \centering
      \caption{Effect of refresh on the performance of OnePerc, considering 4-qubit resource state, a fusion success rate of 0.75, refresh rate of 50 logical layers, and 32GB of RAM.}
      
    \resizebox{0.48\textwidth}{!}{
        \renewcommand*{\arraystretch}{1}
        \begin{footnotesize}
        \begin{tabular}{|p{1.7cm}|p{1.7cm}|p{1.7cm}|p{1.7cm}|}  \hline
        Benchmark & \#Qubits & Non-refreshed \#RSL & Refreshed \#RSL\\ \hline
        \multirow{3}{*}{QAOA} & 25 & 882 & 999 \\  
        \cline{2-4}  & 64 & - & 4,284\\ 

        \cline{2-4} & 100 & - & 8,325 \\
        \hline
         \multirow{3}{*}{QFT} & 25 & 2,271 & 2,637 \\ 
        \cline{2-4}  & 64 & - & 9,945\\
        \cline{2-4} & 100 & - & 19,494 \\
        \hline
        \multirow{3}{*}{RCA} & 25 & 3,252 & 3,870 \\   
        \cline{2-4} & 64 & - & 10,206\\
        \cline{2-4} & 100 & - & 16,056 \\
        \hline  
        \multirow{3}{*}{VQE} & 25 & 759 & 774 \\ 
        \cline{2-4}  & 64 & - & 3,555\\ 

        \cline{2-4} & 100 & - & 7,551 \\
        \hline

        \end{tabular}
        \end{footnotesize}
    }
    \label{tab:refresh}
\end{table}

\subsection{Experiment Result}
In this subsection, we first show the performance of our compiler in comparison with OneQ, then analyze the effects of underlying resource states, hardware size and fusion success probability, for which we only focus on the \#RSL metric. 
This is because unlike OneQ, the \#fusion in OnePerc is predictable from its \#RSL, thus following a same trend with \#RSL.
\paragraph{Performance}
Table~\ref{tab:evaluation} presents the comparison of our framework with OneQ.
The results indicate a significant reduction of \#RSL by our framework, as well as a significant reduction of \#fusion when the circuits are beyond 4 qubits. Specifically, the experiments show that OneQ can work only in the region of small programs and high fusion success probabilities. When the fusion success probability decreases to a practical value around 0.75, it takes more than $10^6$ RSLs to even execute the 4-qubit benckmarks. This implies OneQ's non-scalability due to its lack of capability in systematically handling the randomness of fusion failure. 
In contrast, our framework can work well with a practical success probability, demonstrating an increasing outperformance over OneQ as the programs scale up.

An obstacle of scaling up the experiments in Table~\ref{tab:evaluation} is the large classical memory required in the real-time stage for the storage of graph information. Indeed, the 64-qubit benchmarks in Table~\ref{tab:evaluation} takes a RAM as much as 192 GB. This can be overcome by the refresh mechanism proposed in Section~\ref{sect:tech3}, with an overhead of increased \#RSL. Under the practical fusion success rate of 0.75, Table~\ref{tab:refresh} shows the effect of refresh given 32 GB RAM. It can be seen that while the 32 GB RAM can only afford 25-qubit benchmarks without refresh, it allows for benchmarks of up to 100 qubits with a refresh every 50 logical layers. Compared with the performance of 25-qubit benchmarks (Table~\ref{tab:evaluation} or \ref{tab:refresh}) and 64-qubit benchmarks (Table~\ref{tab:evaluation}) without refresh, the introduction of refresh leads to an average increase of 15.6\% in \#RSL for 25-qubit benchmarks and an average increase of 13.3\% in \#RSL for 64-qubit benchmarks.

\subsection{Sensitivity Analysis}
\begin{figure*}[tp]
        \centering
        \includegraphics[width=0.33\linewidth]{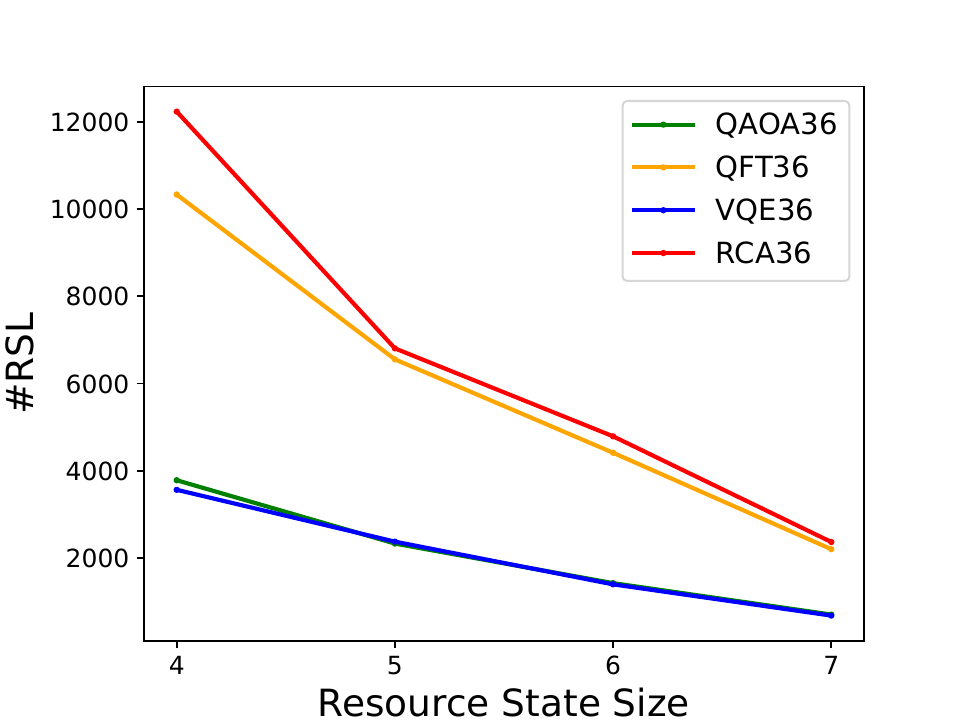}
        \includegraphics[width=0.33\linewidth]{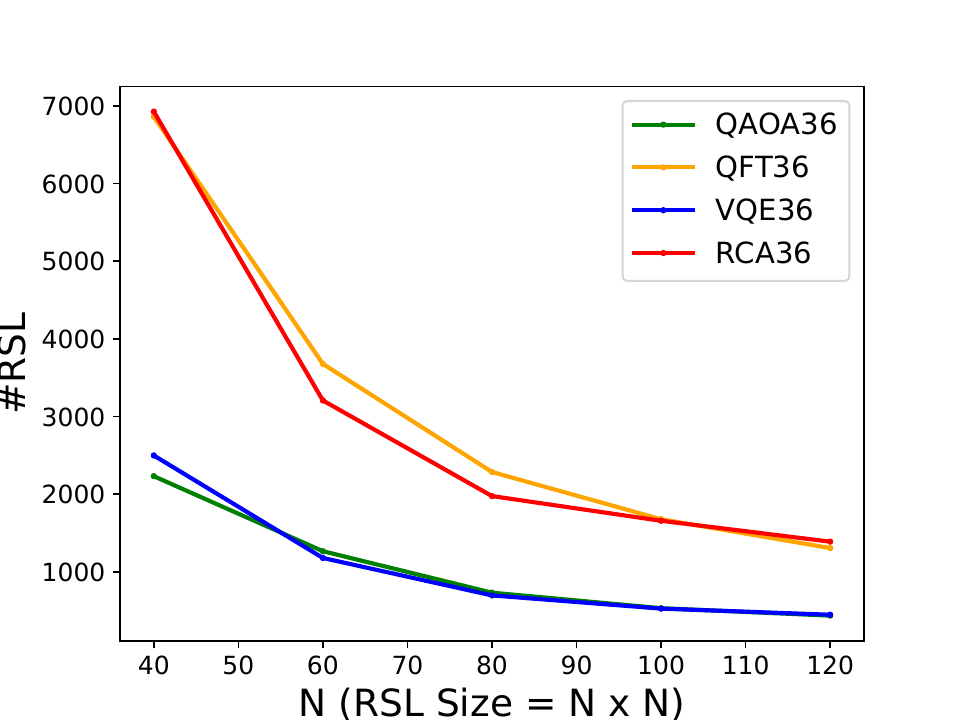}
        \includegraphics[width=0.33\linewidth]{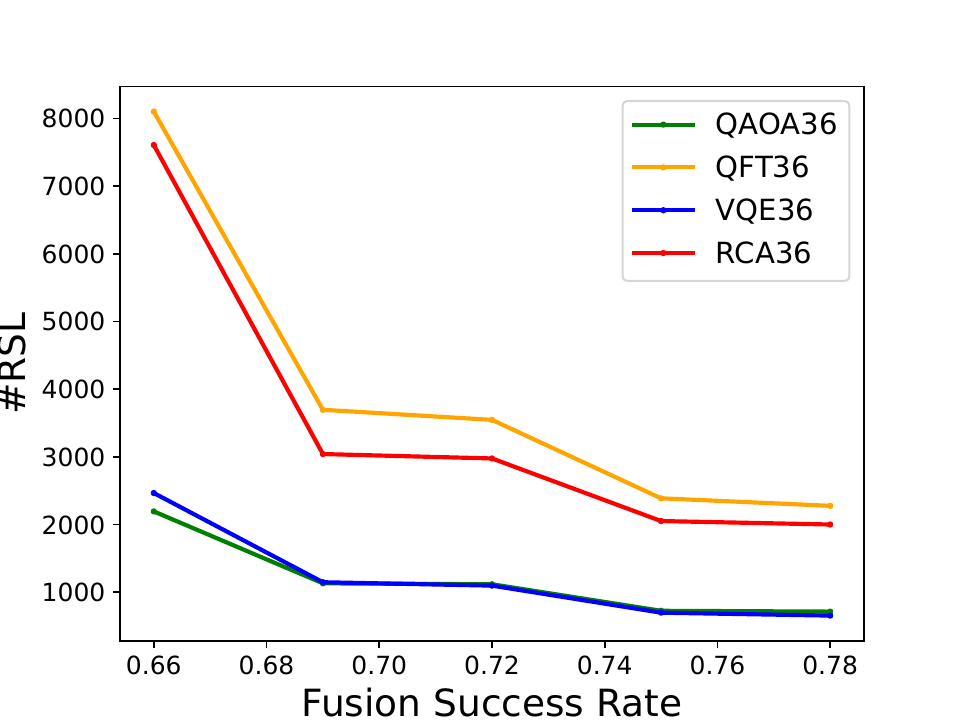} \\
        \hspace{10pt}(a)\hspace{160pt}(b)\hspace{160pt}(c)
        \caption{Effects of resource state size (a), hardware size (b) and fusion success probability (c), with the resource states being 7-qubit ones for (b)(c), hardware size being 84x84 for (a)(c), and the fusion success probability being 0.75 for (a)(b).}
        
        \label{fig:sensitivity}
\end{figure*}

\paragraph{Resource State Size} 
Our compiler has a general applicability to the underlying resource states of various sizes. Fig.~\ref{fig:sensitivity}(a) illustrates the varying \#RSL when executing the programs with star-like resource states of different sizes, i.e., consisting of different numbers of photonic qubits. It can be seen that the \#RSL decreases as the size of resource states increases. This is because a larger resource state can participate in fusions with more qubit degrees, without the need of increasing the degrees by merging multiple RSLs. 

\paragraph{Hardware Size} 
Our compiler has an adaptability to various hardware sizes. Fig.~\ref{fig:sensitivity}(b) shows the varying \#RSL when executing the programs on photonic hardware of different RSL sizes. It can be seen that a larger photonic hardware leads to a reduced \#RSL, which indicates that our framework can effectively utilize the computing resource as it scales up. In particular, a larger RSL can enable a larger renormalized lattice, thus a larger virtual hardware. 
This provides the offline mapping with an increased space for flexible routing, thereby reducing the required logical layer and the \#RSL.

\paragraph{Fusion Success Probability} 
Our compiler has a capability of tolerating fusion failures at a practical level. Fig.~\ref{fig:sensitivity}(c) shows the varying \#RSL when executing the programs under different fusion success probabilities. It can be seen that our compiler can tolerate a fusion success probability as low as 0.66, with the \#RSL decreasing as the fusion success probability increases. 
This is because a higher fusion success probability results in a larger renormalized lattice on RSLs, enabling a larger virtual hardware. This provides the offline mapping with an increased space for flexible routing,
thereby reducing the required logical layer and the \#RSL.

\subsection{Scalability}

\begin{figure*}[tp]
        \centering
        \includegraphics[width=0.33\linewidth]{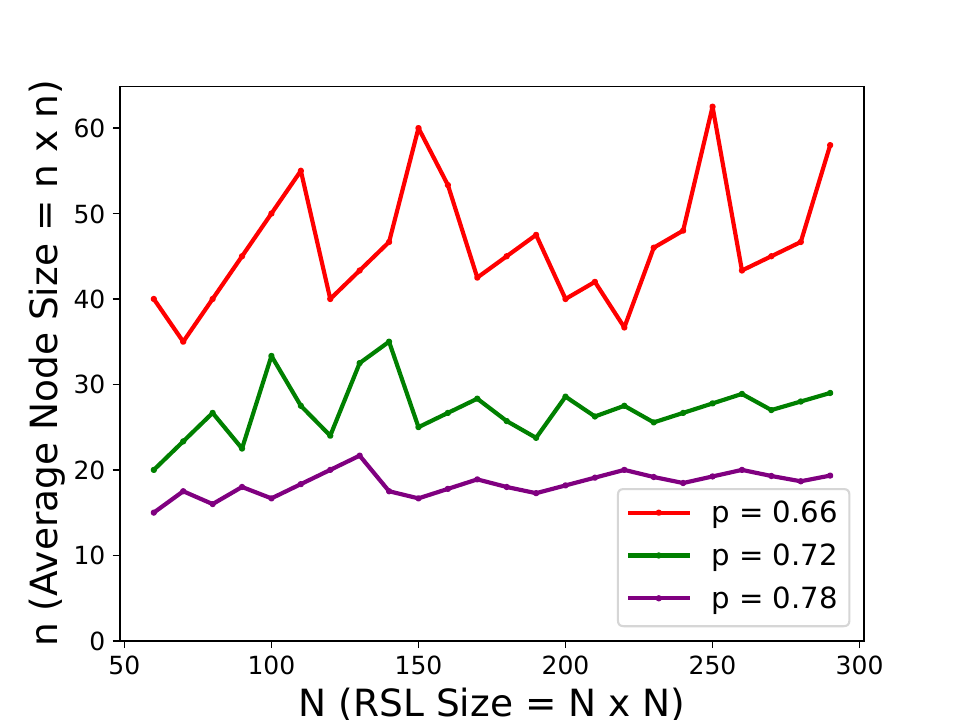}
        \includegraphics[width=0.33\linewidth]{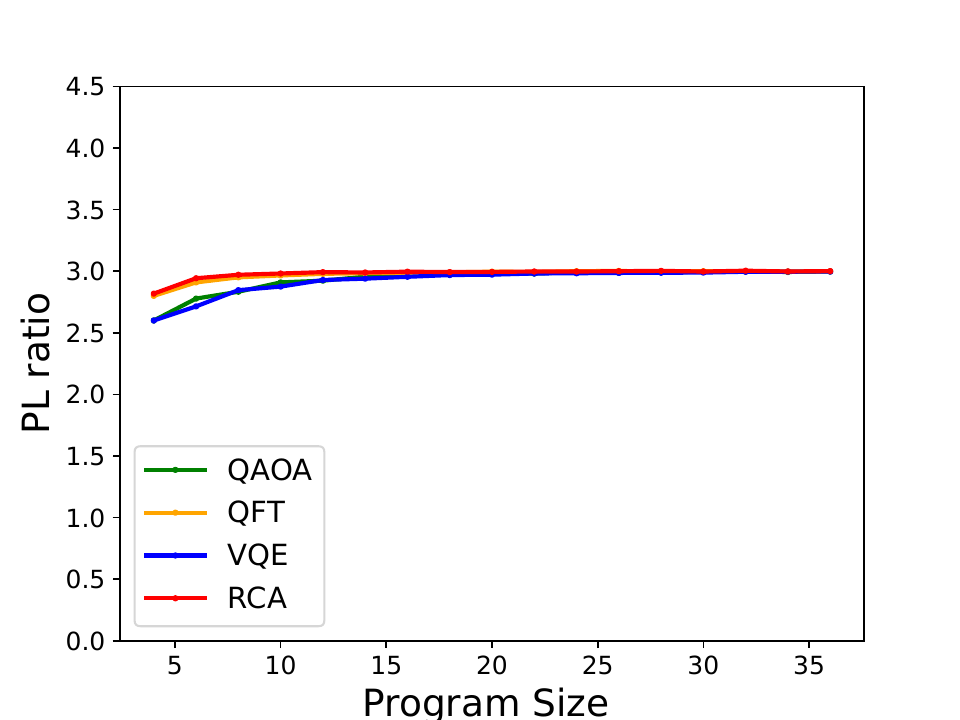} 
        \includegraphics[width=0.33\linewidth]{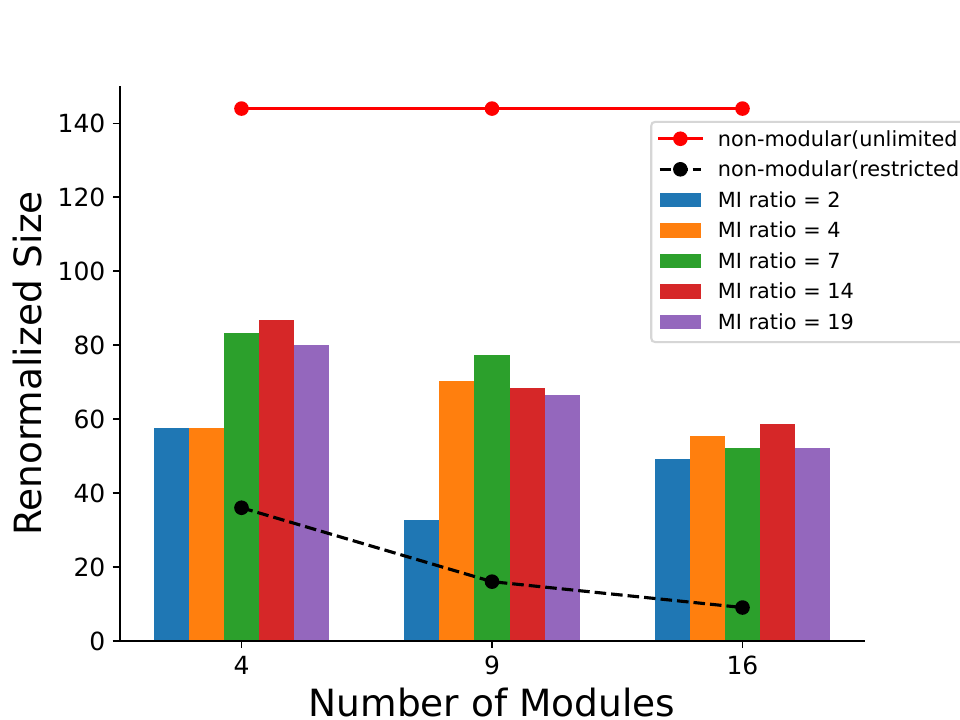} \\
        (a)\hspace{160pt}(b)\hspace{150pt}(c)
        \caption{Scalability and parallelism of OnePerc with 7-qubit resource states.
        The node size $n\times n$ in (a) corresponds to the smallest node size where the renormalization success rate approaches to 1 in Fig.~
        \ref{fig:scalability}.}
        
        \label{fig:scalability1}
\end{figure*}

\paragraph{Resource Consumption} 
Our compiler presents a great scalability in resource consumption, characterized by the stable overhead as the computing scales up.
Fig.~\ref{fig:scalability1}(a) shows the suitable average node size of 2D renormalization as the hardware size increases, corresponding to the average node size at which the renormalization success probability approaches 1 in Fig.~\ref{fig:scalability}. As can be seen, it keeps stable against the variation of hardware size, being smaller with a higher fusion success probability.
Fig.~\ref{fig:scalability1}(b) shows the average ratio of RSL to logical layers as the program size increases. It first increases with the program size and then soon gets stable at a value around 3, implying the successful formation of a logical layer about every 3 RSLs. 
These stable behaviours provide a predictability of the resource consumption and ensures the scalability of our framework.

\paragraph{Modularity Overhead} 
The real-time scalability of our framework can be greatly enhanced through a modular 2D renormalization, which reduces the latency for each RSL by a factor corresponding to the modular number. 
However, this comes with an overhead,
as the presence of intervals between the modules (as illustrated in Fig.~\ref{fig:modular_renormalization}) reduces the available resource on each RSL. To evaluate this resource overhead, Fig.~\ref{fig:scalability1}(c) depicts the size of the renormalized 2D lattice against the number of modules, with the MI ratio (as defined in Fig.~\ref{fig:modular_renormalization}) ranging from 2 to 19. For comparison, the red dots represent the renormalized 2D lattice size by a non-modular algorithm in an unlimited time, while the black dots represent the renormalized size by the non-modular algorithm in a time restricted by that consumed by the modular approach.

\begin{figure}[h!]
        \centering
        
        \includegraphics[width=0.78\linewidth]{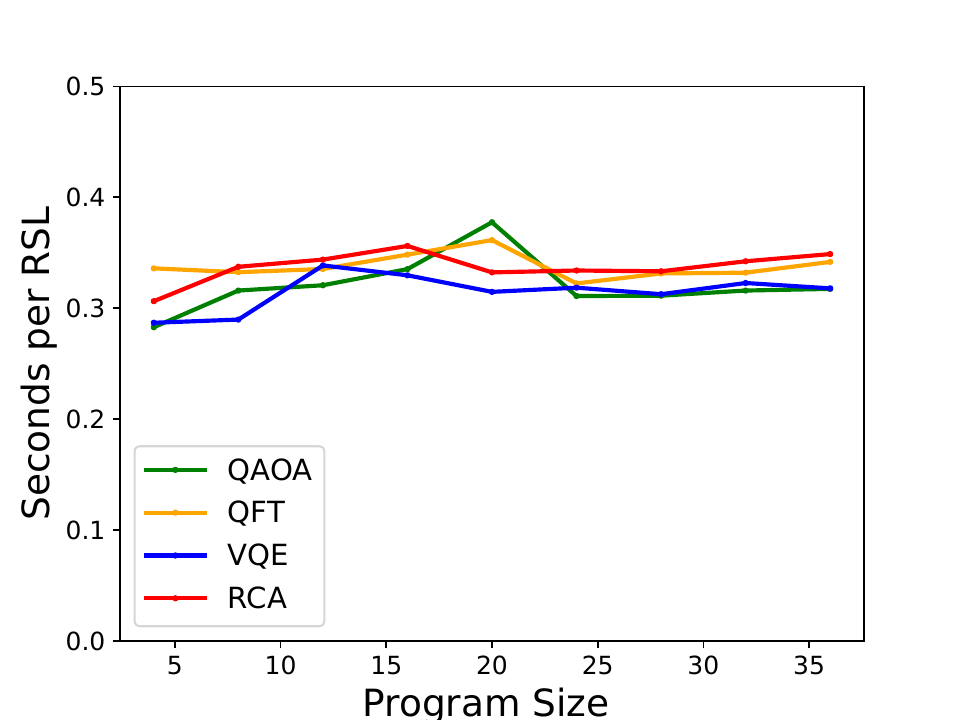}\\
        (a)

        \includegraphics[width=0.78\linewidth]{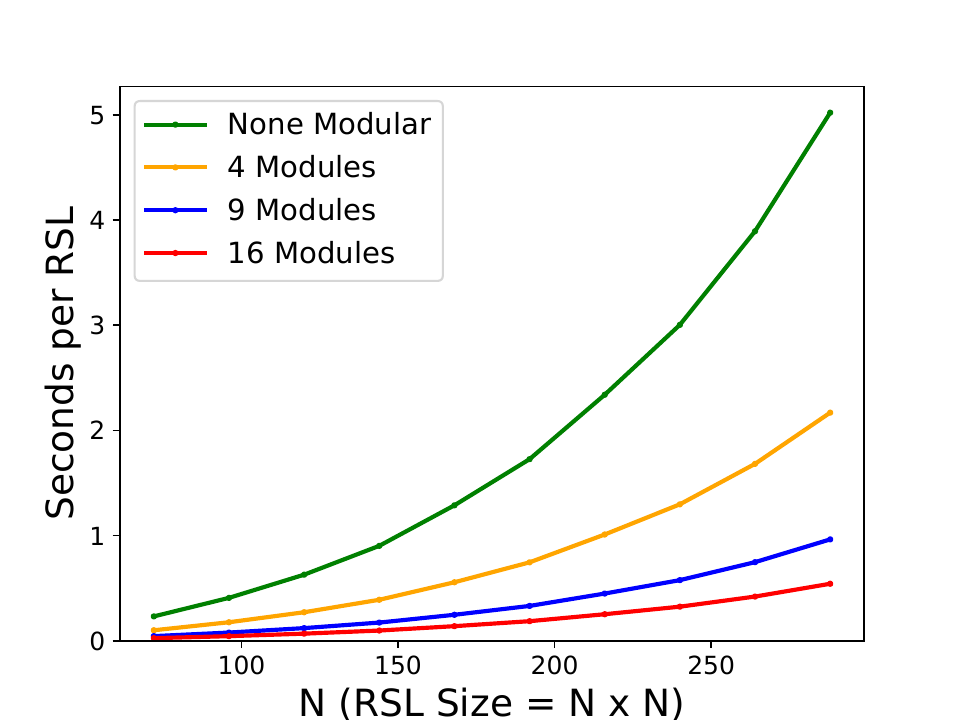}\\
        (b)
        \caption{Oneline proccessing time for each RSL with 7-qubit resource states.  RSL size is $96\times 96$ for (a); fusion success rate is 0.75 for (a)(b);  average node size is chosen as $24\times 24$ for (a)(b); MI ratio is chosen as 7 for (b).}
        
        \label{fig:oneline_compilation_time}
\end{figure}

\begin{figure}[h!]
        \centering
        
        \includegraphics[width=0.78\linewidth]{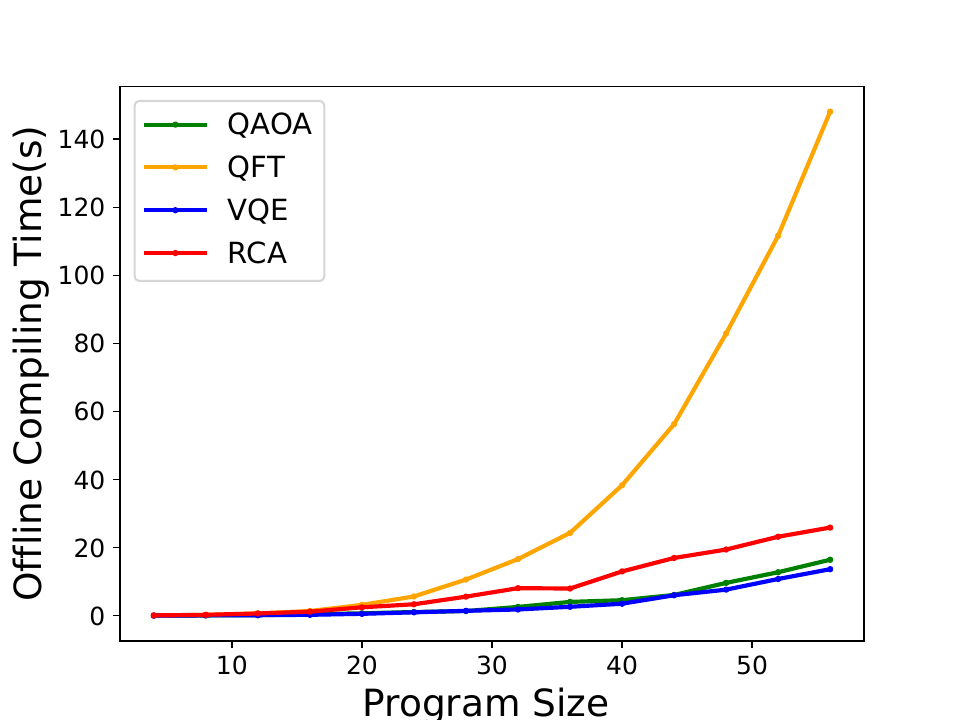}\\
        (a)

        \includegraphics[width=0.78\linewidth]{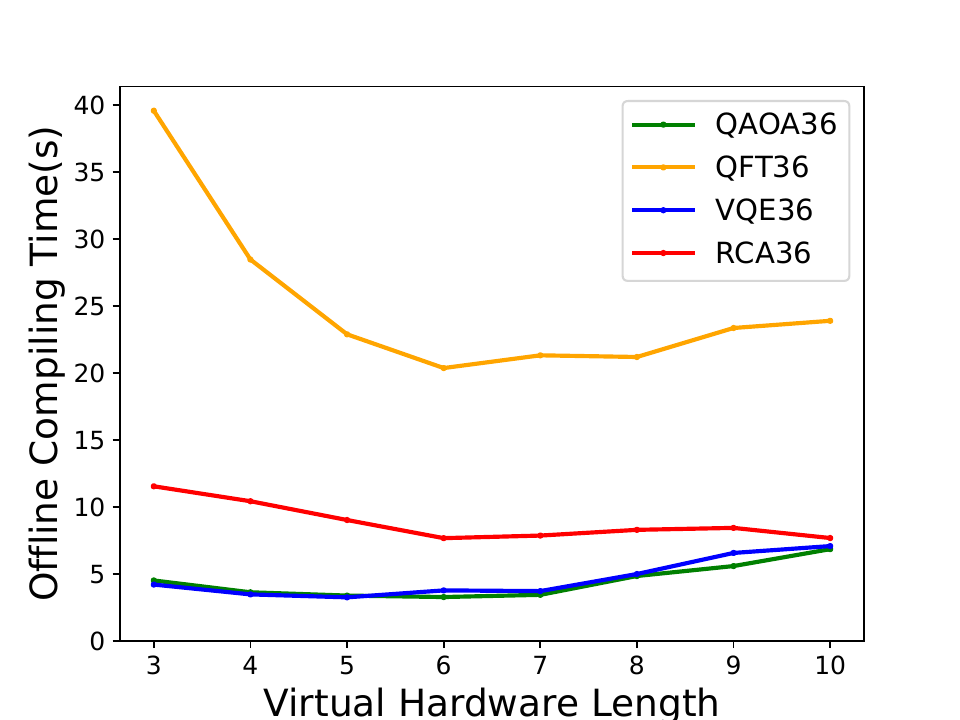}\\
        (b)
        \caption{Offline compilation time on a virtual hardware, with the virtual hardware size being $4\times 4$ for (a). The virtual hardware sizes correspond to the sizes that can be formed by the RSL settings in Fig.~\ref{fig:oneline_compilation_time}. }
        
        \label{fig:offline_compilation_time}
\end{figure}

It can be seen that the size of renormalized 2D lattice by the modular approach is around 60\% of that by the non-modular approach with unlimited time (red), which decreases slightly with the number of modules. This is because an increased number of modules leads to a higher probability of being unable to connect the corresponding paths across different modules. However, the renormalized lattice is significantly larger than that can be achieved by the non-modular approach restricted in the same time (black), ranging from $2\times$ to $6\times$ as the number of modules increases from 4 to 16. This is very important since the time for the online algorithm is always restricted by the limited lifetime of photons. Overall, Fig.~\ref{fig:scalability1}(c) indicates that the modular approach in our framework can significantly improve the real-time scalability with a reasonable overhead of computing resource.

\paragraph{Compilation Time} We show the online and offline compilation time of the benchmarks in Fig.~\ref{fig:oneline_compilation_time} and Fig.~\ref{fig:offline_compilation_time}, with the compiler implemented in Python.
From Fig.~\ref{fig:oneline_compilation_time}(a) it can be seen that the online processing time for each RSL stays stable as the program size increases.
From Fig.~\ref{fig:oneline_compilation_time}(b), which takes an average of all 36-qubit benchmarks,
it can be seen that the processing time for each RSL increases with RSL size, but can be significantly reduced by employing a modular renormalization. For offline compilation time, Fig.~\ref{fig:offline_compilation_time}(a) shows that it increases with the program size. Fig.~\ref{fig:offline_compilation_time}(b) shows that it decreases with the virtual hardware size first and then increases. 
This occurs because an excessively small virtual hardware size leads to a significant total depth, whereas an overly large virtual hardware size results in extended compilation times for each logical layer.

\subsection{Hyper Parameters}
\paragraph{MI Ratio} The sizes of renormalized lattices rely on a suitable choice of MI ratio (defined in Fig.~\ref{fig:modular_renormalization}). Fig.~\ref{fig:scalability1}(c) illustrates the renormalized lattice size with different choices of MI ratios. It can be seen that the renormalization size first increases with the MI ratio and then slightly decreases, peaking at a value around 7.
This is because an excessively low MI ratio leads to a waste of resource with its wide interval space, while an overly high MI ratio increases the probability of unable to connect corresponding paths with its restricted routing space in the intervals.

\paragraph{Average Node Size} A suitable choice of average node size is also important, as it determines the target size of a successful 2D renormalization. Fig.\ref{fig:scalability} illustrates the success probability of reaching different predetermined lattice sizes, i.e., different choices of average node size. It can be seen that the success probability approaches 1 rapidly as the target lattice becomes more coarse-grained. This sharp transition motivates us to choose the smallest average node size that brings the success probability close to 1.

%% file: 08_conclusion.tex
\begin{figure}[h!]
        \centering
        \hspace{-15pt}\includegraphics[width=0.69\linewidth]{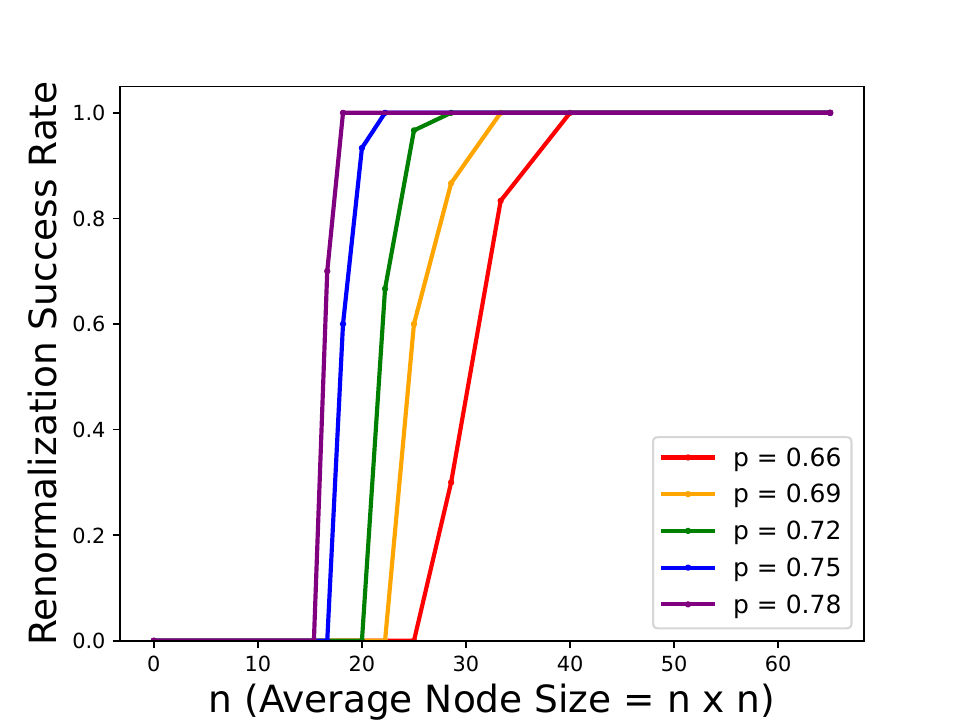}
        \caption{Effect of choices of average node size, with RSL size being $200\times 200$.}
        
        \label{fig:scalability}
\end{figure}
   
\section{Conclusion}
In this work, we provide in-depth analysis and discussion of the challenges for photonic quantum compilation brought by the probabilistic operations involved in the computing. 
We propose a randomness-aware compiler to handle these probabilistic operations, demonstrating a concurrent achievement of scalability and efficiency on photonic systems.
Nevertheless, we believe that there is still significant potential for fully exploring the optimization space. 
We hope that our work could attract more effort from the computer architecture and compiler community to explore the advantages of photonic quantum computing and overcome the unique challenges.

%% file: 10_ackn.tex
\section{Acknowledgement}

We thank the anonymous reviewers for their constructive feedback and the cloud bank~\cite{cloud_credit}.
This work is supported in part by Cisco Research, NSF 2048144 and Robert N.Noyce Trust.

%% file: ae.tex
\fontsize{10pt}{12pt}\selectfont
\appendix
\section{Artifact Appendix}

\subsection{Abstract}

The artifact contains source codes of OnePerc and necessary code scripts to reproduce key results (Table \ref{tab:evaluation}, \ref{tab:refresh}, Fig. \ref{fig:sensitivity},\ref{fig:scalability1},\ref{fig:oneline_compilation_time},\ref{fig:offline_compilation_time},\ref{fig:scalability}) and compare with the baselines in our evaluation. The hardware requirement is a regular X86 server. The software dependencies only contain common python packages. As results in Section~\ref{sect:eval} are averaged over multiple executions, slight deviation is expected in the reproduction.

\subsection{Artifact check-list (meta-information)}

{\small
\begin{itemize}
  \item {\bf Algorithm: }OnePerc contains two major algorithms. 
  \begin{itemize}
    \item The directory Graph\_State\_Mapping/ is dedicated to the offline passes, comprising several essential components:
    \begin{itemize}
        \item Construct\_Test\_Circuit.py creates benchmark circuits with a specific number of qubits.
        \item Graph\_State.py transforms the generated quantum circuits into corresponding program graph states.
        \item Determine\_Dependency.py examines the dependency relationships within the entire graph state.
        \item Mapping\_Routing.py maps the entire graph state onto a virtual hardware of a specific size.
    \end{itemize}

   \item The directory Renormalization/ is dedicated to the online passes, comprising the following key components:
    \begin{itemize}
         \item Percolate.py simulates probabilistic fusion within a real physical scenario to generate a physical graph state.
         \item Renormalization.py reshapes the generated physical graph state to the desired shape of the IR graph state obtained in the offline pass.
        \item Draw\_Grid.py executes 2D renormalization within a single resource state layer.
        \item Check\_Connectivity.py verifies the presence of time-like connections defined in the previous offline pass.
    \end{itemize}
  \end{itemize}

  \item {\bf Output: }The output of the compilation process is the reshaped physical graph state identified within the layers of the physical resource state.
  
  \item {\bf Run-time environment: }Python, Jupyter Notebook.
  \item {\bf Hardware: } Memory size depends on the benchmark size and whether the refresh is enabled (the largest benchmarks without refresh can be processed with 192 GB RAM).
  \item {\bf Experiments: }Compiling the benchmark programs with OnePerc, using OneQ compiler as the baseline.
  \item {\bf Required disk space (approximately): } When selecting the refresh option, only 32GB of disk space is necessary, whereas opting out could necessitate 130GB of disk space.
  \item {\bf Metrics: } Resource state depth and fusion cost.
  \item {\bf Time needed to complete experiments: } The approximate execution time for each benchmark ranges from 10 seconds to 2 hours with program size expanding for OnePerc. For OneQ basline, the execution time can be infinit. In the experiement setting, it is given a upper bound to cost $10^6$ resource state layers, after which the execution time varies from 3 minutes to 6 hours with program size expanding. It will take hundreds of CPU hours to fully reproduce all results in Table \ref{tab:evaluation}, \ref{tab:refresh} and Fig. \ref{fig:sensitivity}, \ref{fig:scalability1}, \ref{fig:oneline_compilation_time}, \ref{fig:offline_compilation_time}, \ref{fig:scalability}.
  \item {\bf Publicly available: }Yes
  \item {\bf Code licenses: } Apache License 2.0
  \item {\bf Workflow framework used: } Jupyter, Qiskit, PyZX
  \item {\bf Archived repo: } https://zenodo.org/records/10799879
  \item {\bf DOI: } 10.5281/zenodo.10799879
\end{itemize}
}

\subsection{Description}

\subsubsection{How to access}

This artifact can be downloaded at the link
\url{https://zenodo.org/records/10799879}.

\subsubsection{Hardware dependencies}
A standard server with Intel CPUs can effectively run our artifact, with the capacity of RAM potentially constraining the scale of benchmarks that can be executed. In our experiments, we allocated 192GB of RAM to accommodate the execution of all benchmarks. However, activating the refresh option would reduce this requirement to just 32GB of RAM.

\subsubsection{Software dependencies}
The artifact is developed using Python 3.10, and we require Jupyter Notebook for its utilization. We have prepared files containing scripts to facilitate the automatic and interactive reproduction of results for convenient validation. Pyzx is employed for generating specific quantum circuits, while other dependencies such as NetworkX, Matplotlib, and NumPy are also utilized.

\subsection{Installation}

To use our artifact, you may download the repo to your local machine from \url{https://zenodo.org/records/10799879} and install the software dependencies by running commands:

\vspace{-10pt}
\begin{align*}
&\texttt{
    conda create -n oneperc python=3.10}\\
&\texttt{
    pip3 install -r requirements.txt
    }
\end{align*}

\subsection{Evaluation and expected results}

After downloading the artifact and installing all software dependencies, you can open the following jupyter notebook files to reproduce experimental data of baseline and OnePerc for corresponding table and figures.

\begin{itemize}
    \item Compiler.ipynb (Table \ref{tab:evaluation})
    \item refesh.ipynb (Table \ref{tab:refresh})
    \item sensitivity.ipynb (Fig. \ref{fig:sensitivity})
    \item scalability.ipynb (Fig. \ref{fig:scalability1}, \ref{fig:scalability})
    \item time.ipynb (Fig. \ref{fig:oneline_compilation_time}, \ref{fig:offline_compilation_time})
    
\end{itemize}

The previous experimental data has already been saved to data/. In scalability.ipynb, sensitivity.ipynb, time.ipynb and refresh.ipynb,
setting `\texttt{RunAgain = False}' will generate the plots directly from original data, while setting `\texttt{RunAgain = True}' will run the experiments again, generating new data and new plots. Note that running with a parameter $N$ corresponds to an $N^2$-qubit benchmark instead of $N$-qubit.

The generation of Table~\ref{tab:evaluation} is the most time-consuming procedure in the evaluation. This is because OneQ performs badly for large programs. Although we force the compilation to terminate when the consumed \#RSL reaches $10^6$, it can take hours to reach this limit. As a result, we provide three code blocks in Compiler.ipynb. 
\begin{itemize}
    \item The first code block allows users to run OneQ for individual benchmarks. By changing the value of $N$, users can obtain the result of OneQ for an $N^2$-qubit benchmark. We recommend users to try benchmarks from small to large and feel free to stop at the scale they obtain a $10^6$ \#RSL for OneQ since larger scales should also lead to a $10^6$ \#RSL.
    \item The second code block allows users to run OnePerc for individual benchmarks. By changing the value of $N$, users can obtain the result of OnePerc for an $N^2$-qubit benchmark. In this process, users can monitor the consumed RAM manually in the terminal.
    \item The third code block allows users to obtain all results of OnePerc in Table~\ref{tab:evaluation} in one shot.
\end{itemize}
The experiment results of these code blocks will be automatically saved to data/. After the experiments, users can run Compiler\_Table.ipynb to read the saved data and generate Table~\ref{tab:evaluation}.

Table~\ref{tab:refresh} can be generated by refresh.ipynb, which runs the offline mapping and obtain an estimated \#RSL from the number of logical layer. This is because from Fig.~\ref{fig:scalability1}(b), we know that \#RSL has a stable relation with the number of logical layers.
The results of non-refreshed \#RSL for 25-qubit benchmarks can be obtained directly from Table~\ref{tab:evaluation}.
The `-' in Table~\ref{tab:refresh} means that the compilation utilizes more than 32 GB RAM. 
In the execution of OnePerc for individual benchmarks (code block 2 in Compiler.ipynb), users can monitor the consumed RAM manually in the terminal (e.g., using \texttt{htop} on Linux). 
When running the compiler with large enough RAM, users will observe that the consumed RAM exceeds 32 GB for benchmarks larger than 25 qubits. 
When running the compiler with only 32 GB RAM, users will observe that the compilation of benchmarks larger than 25 qubits would be killed after some time.